%% file: main.tex
\documentclass[acmsmall]{acmart}

\AtBeginDocument{%
  \providecommand\BibTeX{{%
    \normalfont B\kern-0.5em{\scshape i\kern-0.25em b}\kern-0.8em\TeX}}}
    
\setcopyright{acmlicensed}
\acmJournal{PACMHCI}
\acmYear{2024} \acmVolume{8} \acmNumber{CSCW2} \acmArticle{470} \acmMonth{11}\acmDOI{10.1145/3687009}



\usepackage{graphicx} % Required for inserting image

\title[``Come to us first'': Centering Community Organizations in Artificial Intelligence for Social Good Partnerships]{``Come to us first'': Centering Community Organizations in Artificial Intelligence for Social Good Partnerships}

\author{Hongjin Lin}
\orcid{0000-0001-6207-2147}
\affiliation{%
  \department{School of Engineering and Applied Sciences}
  \institution{Harvard University}
  \city{Allston}
  \state{Massachusetts}
  \country{United States}}
\email{hongjin_lin@g.harvard.edu}

\author{Naveena Karusala}
\orcid{0000-0003-0355-3041}
\affiliation{%
  \department{School of Engineering and Applied Sciences}
  \institution{Harvard University}
  \city{Allston}
  \state{Massachusetts}
  \country{United States}}
\email{naveenak@seas.harvard.edu}

\author{Chinasa T. Okolo}
\orcid{0000-0002-6474-3378}
\affiliation{%
  \department{Center for Technology Innovation}
  \institution{The Brookings Institution}
  \city{D.C.}
  \state{Washington}
  \country{United States}}
\email{cokolo@brookings.edu}

\author{Catherine D’Ignazio}
\orcid{0000-0002-8673-1941}
\affiliation{%
  \department{Urban Studies and Planning}
  \institution{MIT}
  \city{Cambridge}
  \state{Massachusetts}
  \country{United States}}
\email{dignazio@mit.edu}

\author{Krzysztof Z. Gajos}
\orcid{0000-0002-1897-9048}
\affiliation{%
  \department{School of Engineering and Applied Sciences}
  \institution{Harvard University}
  \city{Allston}
  \state{Massachusetts}
  \country{United States}}
\email{kgajos@seas.harvard.edu}

\renewcommand{\shortauthors}{Hongjin Lin et al.}

\date{\today}

\begin{document}

\begin{abstract}
% the context
Artificial Intelligence for Social Good (AI4SG) has emerged as a growing body of research and practice exploring the potential of AI technologies to tackle social issues. This area emphasizes interdisciplinary partnerships with community organizations, such as non-profits and government agencies. However, amidst excitement about new advances in AI and their potential impact, the needs, expectations, and aspirations of these community organizations--and whether they are being met--are not well understood. Understanding these factors is important to ensure that the considerable efforts by AI teams and community organizations can actually achieve the positive social impact they strive for. Drawing on the \textit{Data Feminism} framework, we explored the perspectives of community organization members on their partnerships with AI teams through 16 semi-structured interviews. Our study highlights the pervasive influence of funding agendas and the optimism surrounding AI's potential. Despite the significant intellectual contributions and labor provided by community organization members, their goals were frequently sidelined in favor of other stakeholders, including AI teams. While many community organization members expected tangible project deployment, only two out of 14 projects we studied reached the deployment stage. However, community organization members sustained their belief in the \textit{potential} of the projects, still seeing diminished goals as valuable. To enhance the efficacy of future collaborations, our participants shared their aspirations for success, calling for co-leadership starting from the early stages of projects. We propose \textit{data co-liberation} as a grounding principle for approaching AI4SG moving forward, positing that community organizations’ co-leadership is essential for fostering more effective, sustainable, and ethical development of AI.

% the problem
% our approach
% our findings
% contributions

\end{abstract}

\maketitle

\input{1-Introduction}

\input{2-Background}
\input{3-Methodology}

\input{4-Findings}

\input{5-Discussion}

\input{6-Conclusion}

\begin{acks}
We are thankful to all participants who contributed valuable insights to our study. We are grateful to the members of the \textit{Intelligent Interactive Systems Group} at Harvard and the \textit{Data + Feminism Lab} at MIT for their valuable feedback on the early drafts of this work. 

This work was partly supported by the National Science Foundation under Grant No. IIS-2107391. Any opinions, findings, conclusions, or recommendations expressed in this material are those of the author(s) and do not necessarily reflect the views of the National Science Foundation.
\end{acks}

\bibliographystyle{ACM-Reference-Format}
\bibliography{2024CSCW-CommAI4SG}

\received{January 2024}
\received[revised]{April 2024}
\received[accepted]{May 2024}

\end{document}

%% file: 1-Introduction.tex
\section{Introduction}

% 1: context
Artificial Intelligence for Social Good (AI4SG) has emerged as a growing body of research and practice that explores the potential of AI technologies for tackling complex social issues such as climate change, humanitarian aid, and public health concerns~\cite{shi_artificial_2020, perrault_ai_2020, tomasev_ai_2020, cowls_ai_2021, li_ai_2021, aula_stepping_2023}. Often, AI4SG initiatives rely on interdisciplinary partnerships with community organizations, such as non-profits and government services, who are domain experts on the social issues AI is being applied to~\cite{shi_artificial_2020, perrault_ai_2020, berendt_ai_2019, tanweer_cross-sector_2017}. Through academic publications, both industry-based and academic technology teams demonstrate (implicitly or explicitly) the tangible benefits they receive from such partnerships: well-motivated problems that translate into novel computational challenges, access to data, and the ability to claim potential real-world impact (e.g.,~\cite{shi_artificial_2020}). Ultimately though, the stated goals of these partnerships and the research produced through them are typically to support community organizations and the positive change they enable in the world. Understanding the full impact of these partnerships, thus, requires answering the following questions: What are the motivations, needs, and aspirations of collaborators within community organizations, and are they being met? Pursuing these questions is necessary to ensure that the efforts of both technology teams and community organizations actually achieve the social good outcomes they strive for.

% 2: HCI scholarship and state of research
Computer-Supported Cooperative Work (CSCW), Human-Computer Interaction (HCI), and Information and Communication Technologies and Development (ICTD) researchers have studied meaningful engagement with community organizations in technology design and implementation, and there has been an increasing focus on equitable relations in the production of AI and data-driven technologies specifically. Prior work has described the power asymmetries in collaborations around the development and implementation of AI technologies~\cite{ismail_ai_2021, crawford_atlas_2021, espinoza_big_2021, magalhaes_giving_2021, dignazio_data_2020}. These power asymmetries result in deskilling of field workers and domain experts by AI developers~\cite{sambasivan_deskilling_2022}, AI occupying an unquestionable ``authority'' status among users~\cite{kapania_because_2022, okolo_it_2021}, erosion of mission-driven organizations' autonomy~\cite{bopp_disempowered_2017}, and other negative impacts on stakeholders. We hone in on community organizations, who are often key mediators between technologists, field workers, and end beneficiaries, while also being especially vulnerable to power asymmetries due to their dependency on philanthropic grants and external technical expertise~\cite{bopp_disempowered_2017, erete_storytelling_2016, Sum_2023}. While community organizations' internal data practices have been explored in depth (e.g.,~\cite{darian2023enacting, bopp_disempowered_2017, Dell2015paper}), \citet{susha_data_2019} note the need to understand the new external collaborations arising around data-driven technologies for social good. In this paper, we seek to shed light on these collaborations, with the goal of centering community organizations in what equitable production of AI and data-driven technologies could look like.

We anchor our work in D'Ignazio and Klein's \textit{Data Feminism} framework~\cite{dignazio_data_2020} to examine power asymmetries in AI4SG partnerships and foreground the perspectives of collaborators in community organizations. The \textit{Data Feminism} lens inspired us to ask who contributes to AI4SG projects, what forms of knowledge are privileged, and whose goals are prioritized. Specifically, we ask the following research questions: 

\begin{itemize}
    \item \textbf{RQ1:} What are the goals and motivations of community organizations and their members when participating in AI4SG projects? What factors influence their participation?
    \item \textbf{RQ2:} How do community organization members contribute to AI4SG partnerships?
    \item \textbf{RQ3:} How do the outcomes and success metrics of AI4SG projects reflect (or not) community organizations' and other stakeholders' goals? 
\end{itemize}

We addressed these questions by conducting 16 semi-structured interviews with individuals who currently work or have worked in community organizations on 14 AI4SG projects (either completed or ongoing). The community organizations are non-profits, international organizations, and government services. The AI4SG project domains span across humanitarian aid, agriculture, health, misinformation, and more. The community organizations partnered with academic institutions, fellowship programs, and private corporations, whom we call ``technology teams'' in this paper.

Our analysis revealed a pervasive influence of funders' agendas and belief in the promise of AI on collaborations, impacting the motivations, expectations, and priorities of community organizations and their members. Despite the significant intellectual contributions and labor from community organization members, their goals were frequently sidelined in favor of funders' and technology teams' priorities. While many expected a tangible product or deployment from the partnerships, only two out of 14 projects we studied reached the deployment stage. However, when projects fell short of expectations, community organization members sustained their belief in the \textit{potential} of the projects, still seeing diminished goals as valuable. To enhance the efficacy of future collaborations, our participants shared aspirations for future collaborations: 1) adopting a relationship-first approach by involving community organizations in the early ideation stage, 2) allowing community organizations to co-lead throughout the project, and 3) investing in technical capacity building for end users of the AI systems.

% 6: discussion and contribution
By foregrounding the perspectives of community organization members, this work offers an empirical understanding of the power asymmetries within AI4SG projects and where these power differentials manifest. Drawing on our findings, we examine the wider social conditions that generate increasing interest in AI4SG and call for stakeholders, especially funders and technology teams, to shift focus from the tool (AI) to the social issues at hand. Lastly, we propose that technology teams use \textit{data co-liberation}~\cite{dignazio_data_2020} as a grounding principle for AI4SG funding and collaborations moving forward, and we offer recommendations to enact the principle. Taken together, we urge all AI4SG stakeholders to center and properly compensate community organizations for their expertise, assets, and leadership, to support effective, inclusive, and ethical AI development.

%% file: 2-Background.tex
\section{Related Work}

To situate our work, we draw on literature in CSCW, HCI, and ICTD on power dynamics in the production of AI and data-driven systems intended for positive social impact. We also draw on prior work on the relationship between social sector organizations, data, and data scientists. We then describe the grounding theoretical framework \textit{Data Feminism}~\cite{dignazio_data_2020} and how we apply the lens in our work. To this body of work, we contribute insights into the motivations, needs, experiences, and aspirations of community organization members involved in AI4SG partnerships.

\subsection{Power Dynamics in Community-Based Research and AI Development}

Prior research and workshops have reflected extensively on approaches to technology design and implementation when working with communities outside of academic labs and industry settings. Areas of work include participatory design (e.g.~\cite{thinyane2018critical,till2022community, Bratteteig_Wagner_2016, Vines2013configuring, Zytko_2022}), action research (e.g.~\cite{Hayes_2011, LeDantec_2016}), and community-based research (e.g.~\cite{Wan_2023, Liang_2023, dearden2018minimum,brown2019some}). This work aims to engage impacted communities in the design and research process, through various degrees of collaboration and community leadership~\cite{Racadio_2014, Unertl_2016, delgado2023participatory}, but it also emphasizes tensions and power dynamics that can arise in the process~\cite{Lodato2018institutional,Walsham2017ictd, Harris2016how, schelenz_information_2022, Bratteteig2016unpacking, Muhammad2015reflections}. %, highlighting power differentials and strained relationships between researchers and community members or practitioners. For example, %through autoethnography and intersectional analysis of power, 
For example, \citet{erete_method_2023} emphasize that researchers have more power in securing funding, designing methods, and disseminating research outputs in community-based research projects. Recent work calls for contending with the historical context of power~\cite{Harrington_2019, erete_method_2023}, letting go of control and rigid study structures~\cite{LeDantec_Fox_2015, Winschiers-Theophilus_2010, Wallerstein_Duran_2006}, and finding ways to meaningfully connect research and practice towards ``informed practice''~\cite{Kumar2018towards}.

AI technology design introduces unique issues that further complicate community engagement. AI4SG projects often require intensive data work~\cite{thakkar2022machine, crawford_atlas_2021}, as well as careful consideration of differences in expertise and issues of transparency, especially in Global South contexts where many AI4SG projects situate~\cite{Okolo_2024, Okolo2024you} and where data may be more scarce. Prior work has started to examine power dynamics and uneven distribution of benefits among stakeholders in the development of AI systems intended for social impact, as well as in user-centered applications more generally. Stakeholders in such projects that have been studied include frontline workers with domain expertise, dedicated data workers, and AI developers. 

Research has highlighted the impact of misaligned incentives and constraints on frontline workers, which contribute to AI developers blaming workers for poor quality data and devaluing their range of contributions to AI development~\cite{sambasivan_deskilling_2022,thakkar2022machine}. Scholars call for crediting frontline workers' contributions to AI systems~\cite{sambasivan_deskilling_2022}, better supporting workers in enabling data quality~\cite{batool2021detecting}, and developing shared values and goals among stakeholders ~\cite{thakkar2022machine, ismail_public_2023}. \citet{thakkar2022machine} also suggest that accountability for data quality be shared between AI developers and the community organizations providing datasets (rather than individual workers), indicating the key role community organizations can play in enabling ethical relations. 

Other studies focus on the power dynamics imposed on data workers employed by companies that generate and annotate datasets. Prior work has established how directives given to workers and performance metrics prioritize managers' and data requesters' worldviews, and precarious working conditions limit workers' agency in interpreting data~\cite{miceli2022data,miceli2020between}. Meanwhile, data annotators do not get many opportunities for skilling~\cite{wang2022whose}. Demonstrating how making power dynamics visible can reshape AI development, \citet{miceli2022studying} argue that metrics for dataset quality should account for power dynamics that affect the interpretation of data, while \citet{kapania_because_2022} suggest ways of preserving diversity in the data annotation process.  

Prior work has focused on AI developers themselves, particularly in industry settings, and the %Studies have established how their values and practices affect seemingly objective data, models, and success metrics~\cite{muller2019data,passi2020making,sambasivan_everyone_2021}. 
complexities they face in engaging end users. Studies have demonstrated how they require more support in proactively engaging stakeholder needs~\cite{bingley2023human}, especially early in the ideation process~\cite{yildirim2023investigating}. In Global South contexts, access to scarce data requires especially careful navigation of community partnerships on account of the labor required to produce it~\cite{sambasivan_everyone_2021}. When AI developers do engage with stakeholders, prior work has found a reliance on brief consultation over involving stakeholders in decision-making~\cite{delgado2023participatory, Okolo2024you}, and engaging client organizations as proxies more than actual end users~\cite{hartikainen2022human}. 
Toolkits, frameworks, and auditing mechanisms have been created to help AI developers responsibly build tools, but studies have found that factors such as conflicting values within teams \cite{varanasi2023currently}, profit motives \cite{deng2023understanding}, lack of context-specific guidance \cite{heger2022understanding,holstein2019improving,deng2022exploring, wong2023seeing}, or lack of prior experience \cite{balayn2023fairness} shape whether developers can successfully implement them. 
%While less focused on stakeholder engagement, a number of studies have looked at how AI developers implement responsible AI values~\cite{varanasi2023currently}, toolkits~\cite{deng2022exploring, Okolo2024you}, and frameworks in their work~\cite{heger2022understanding}. These studies often find that organizational factors such as conflicting values within teams, profit motives, or lack of prior experience shape whether developers can successfully implement them~\cite{deng2023understanding,balayn2023fairness,heger2020all, varanasi2023currently, Okolo2024you}.  
%Studies also suggest that toolkits and documentation frameworks need to be more use- and context-specific~\cite{deng2022exploring,heger2022understanding,holstein2019improving}, provide more explicit prompts for reflection and critical thinking~\cite{heger2022understanding,balayn2023fairness}, support cross-team collaboration and organizational buy-in~\cite{deng2022exploring,yildirim2023investigating}, and help developers navigate labor and organizational power dynamics~\cite{wong2023seeing}. 

%Some studies propose that toolkits aimed at supporting the development of human-centered AI need to better help navigate more overarching labor, organizational, and institutional power dynamics when implementing AI ethics~\cite{wong2023seeing}, and prompt critical thinking over checkbox culture~\cite{balayn2023fairness}.

The literature has also brought attention to larger beliefs around AI that contribute to its overstated authority and contributions~\cite{aula_stepping_2023, crawford_atlas_2021, eubanks_automating_2018, berendt_ai_2019, okolo_it_2021, Lupetti2024unmaking}. Studies in the Indian context, for example, demonstrate the significant belief in AI authority over human institutions and reduced expectations that AI systems be accountable to users~\cite{kapania_because_2022,ramesh2022platform}. A number of studies have demonstrated how ``data-drivenness'' is sometimes a veneer that invisibilizes the human labor that makes data-driven systems work~\cite{gray_ghost_2019, thakkar2022machine, passi2020making, Dell2015paper}. This has resulted in work that seeks to re-center labor and care in the production and maintenance of AI or data-driven systems~\cite{sun2023care,chen2023maintainers,fox2023patchwork}. Prior work also points out the problematic practices underlying AI4SG initiatives, such as corporations equating social good with data and AI~\cite{magalhaes_giving_2021, aula_stepping_2023,crawford_atlas_2021}, using social good applications to justify extractive data practices~\cite{espinoza_big_2021}, or pushing AI when there are more effective solutions~\cite{holzmeyer2021beyond,radhakrishnan_experiments_2021, pruss_ghosting_2023}. 

We build on this body of work by honing in on community organizations' role in AI4SG projects, and the specific hurdles that AI technologies introduce to community-engaged research in this context. Though other stakeholders, such as end beneficiaries, are also essential, collaborators in community organizations often mediate between AI developers, frontline workers, and end beneficiaries. This makes them crucial actors in the process of making AI development accountable to their overall social mission. By centering community organization members' perspectives on AI4SG collaborations, we shed light on the potential role of their leadership in addressing barriers to ethical AI development. We also expand on how the optimism surrounding AI's potential manifests in partnerships.

\subsection{Social Sector Organizations' Relationships with Data and Data Scientists}

Research has addressed social sector organizations specifically, such as government agencies, non-profits, or activist groups, and their relationship with data and data scientists. Community organizations engaged in AI4SG partnerships fall under this group of social sector organizations, often characterized by a focus on a social mission, resource constraints, and reliance on external expertise~\cite{bopp_disempowered_2017, Kumar2018towards}. 

A body of work has focused on social sector organizations' internal data practices. %, whether they help evaluate programs, make decisions, perform certain values, or support other organizational or political goals \citet{}. %For example, \citet{darian2023enacting} lay out how advocacy organizations use data for amplification of marginalized voices, mobilization of policymakers and funders, legitimizing their own expertise and ideas, and innovation in communicating social issues~\cite{darian2023enacting}. 
Studies have highlighted the unique focus of data work in social sectors compared to other sectors, including the emphasis on care~\cite{zegura2018care}, collective responsibility~\cite{meng2019collaborative}, amplifying marginalized voices~\cite{darian2023enacting}, and mobilizing policymakers and funders~\cite{darian2023enacting}, over surveillance, individualization, or profit. However, prior work has also shown that social sector organizations are not well served by existing infrastructures for data~\cite{darian2023enacting, Dell2015paper,bay2020community, bopp2023}. For example, available open data or big data may not be relevant or accessible~\cite{yoon2020toward}. Power dynamics can also incentivize data collection and use with more powerful stakeholders in mind (such as funders)~\cite{bay2020community,bopp_disempowered_2017, erete_storytelling_2016,darian2023enacting}. %, and the need for information systems that better support the measurement of progress towards complex social impact goals~\cite{bay2020community, bopp2023}. 
Still, prior work notes how, despite limited resources, these organizations ``tinker'' with data~\cite{tran2022careful}, reconfigure databases and data representations~\cite{voida2011homebrew,khovanskaya2020bottom}, and build alliances to achieve their goals~\cite{alvarado2017making}, exemplifying their agency~\cite{tran2022careful,meng2018grassroots}. %They also use data not just to garner resources from power structures, but also to build internal capacity and transform structures~\cite{meng2018grassroots}. 
Prior work emphasizes the importance of education, care, and partnerships to help social sector organizations utilize data for social justice~\cite{sandberg2023re,yoon2020toward}, and other work suggests novel ways that design can support data practices in domains such as human rights~\cite{alvarado2017making,alvarado2023mobilizing}, immigration~\cite{li2018working}, labor organizing~\cite{khovanskaya2020bottom}, food assistance~\cite{boone2023data}, and charities~\cite{elsden2019sorting}. While this work has established the power dynamics that affect data practices, our work contributes an understanding of how power shapes organizations' approach to collaborations around AI and data-driven technologies. 

Prior work has called for greater attention to these new collaborations emerging between public, private, and social sector organizations as a result of the increasing desire to leverage data for social good. \citet{susha_achieving_2023} describes organizations' motivations for sharing data, including the need for resources, responsibility towards their social mission, and the desire to learn from other sectors. Susha et al.'s literature survey \cite{susha_data_2019} notes several challenges that arise in collaborations on data-driven tools, including ambiguous data-sharing policies and ethical guidelines, misaligned incentives and differences in organizational culture, and the complexity of measuring impact and value. %For example, in the domain of environmental sustainability, \citet{espinoza_big_2021} argue that the AI4SG initiatives promote the ongoing control of data by private companies rather than meaningfully addressing environmental issues. %In follow-up work, \citet{susha_achieving_2023} provides a characterization of data-driven social sector partnerships, categorizing organizations' motivations for data-sharing into three models: 1) a resource dependence model featuring organizations' self-interests, 2) a social issue model emphasizing organizations' responsibility for a larger social issue, and 3) a social sector model that motivates organizations to learn from other sectors. \citet{susha_achieving_2023} also advocate for aligned goals, sharing expertise, financial resources, and trust in successful partnerships around data-driven systems. 
How collaborations develop is also dependent on positionality within an organization~\cite{maxwell_data_2016}---actors in social sector organizations themselves may be incentivized to prioritize the values of internal leadership and AI developers over community members~\cite{kawakami_studying_2023}. Studies offer recommendations for reconceptualizing collaborations between social sector organizations and data scientists. Recommendations include building trust, aligning development with existing data practices ~\cite{Jung_2022, alvarado2023mobilizing, Dimas_2023}, and sharing expertise and financial resources \cite{susha_achieving_2023}. Prior work also advocates for acknowledging community organizations' role in translating across stakeholders \cite{Sum_2023,hou2017hacking} and defining complex notions like fairness \cite{ismail_public_2023}, viewing "errors" as an opportunity to re-center overlooked expertise \cite{Lin_Jackson_2023}, and having brokers who can mediate strong collaborations and actionable insights \cite{hou2017hacking,alvarado2023mobilizing}. %Ismail et al. also suggest that social sector organizations can be essential in defining complex notions like fairness and success metrics~\cite{ismail_public_2023}.
%Prior work notes the importance of brokers in mediating such strong collaborations~\cite{hou2017hacking} and delivering actionable insights to non-profits~\cite{alvarado2023mobilizing}. \added{\citet{Sum_2023} emphasize community organizations' role and labor as technology and data translators, mediating between community members and funders to build trust with communities and legitimacy with multiple stakeholders.} 
%\citet{Lin_Jackson_2023} also encourage us to consider how ``errors'' in applied data science collaborations can help create new collaborations or recenter overlooked expertise. %In terms of the use of data, prior work suggests better aligning with available data and drawing on situated data practices to inform system-building, with the aim of making insights more actionable~\cite{Jung_2022, alvarado2023mobilizing, Dimas_2023}. %For example, \citet{Jung_2022} suggest designing tools that can help domain experts define and manipulate the relationship between different variables according to their own situated data practices, which could then inform data scientists' work. 

Expanding upon the existing scholarship, our work contributes a situated understanding of community organization members' perspectives on AI4SG partnerships. By looking at the development of collaborations \textit{over time}, we highlight not just community organization members' motivations to participate in AI4SG, but also how their expectations evolve and their aspirations for future collaborations. This allows us to expand on factors that could better sustain successful collaborations from the perspective of community organization members.

\subsection{Drawing on Data Feminism}

% explain Data Feminism + how others have used it 
Our research is grounded in the conceptual framework of \textit{Data Feminism} presented by \citet{dignazio_data_2020}. \citet{dignazio_data_2020} contend that power within the field of data science is unevenly distributed, highlighting the disproportionate representation of those with technology and engineering proficiency among those engaged in data science tasks. These dominant groups are perceived as experts, and entrusted with decision-making authority over data \cite{dignazio_data_2020}, and this influence can unintentionally exclude alternative viewpoints, especially those of marginalized individuals and groups---a phenomenon termed the ``privilege hazard'' \cite{dignazio_data_2020}. %``Privilege hazard'' hinders the recognition of the lived experiences of marginalized individuals and groups, impeding the development of equitable data products. 
In response to these challenges, \citet{dignazio_data_2020} propose seven guiding principles to illuminate power asymmetries, confront injustices, and orient data science projects toward more inclusive outcomes: 1) Examine Power, 2) Challenge Power, 3) Elevate Emotion and Embodiment, 4) Rethink Binaries and Hierarchies, 5) Embrace Pluralism, 6) Consider Context, and 7) Make Labor Visible. Previous research has applied the principles of \textit{Data Feminism} across diverse domains, including advocacy data work~\cite{darian2023enacting}, data for non-profits~\cite{sandberg2023re}, open data initiatives in disaster relief~\cite{Paudel_Soden_2023}, feminicide data analysis~\cite{Suresh_2022}, textile design~\cite{Lean_2021}, and COVID-19 data in the U.S.~\cite{DIgnazio_Klein_2020}. More broadly, the CSCW and HCI literature has actively embraced feminist theory to analyze technologies and their impact on society (e.g., \cite{Bardzell_Bardzell_2011, Leavy_2021,Ahmed_2020, Bardzell_2010}). %and considered feminism as a design methodology and agenda~\cite{Ahmed_2020, Bardzell_2010}. %\added{~\citet{erete_method_2023} apply Black feminist epistemologies and intersectionality to call for examining the history and the intersecting systems of power, leveraging an autoethnography on the research process of co-designing technology with community outreach workers addressing violence in their predominantly Black neighborhoods.} %Particularly pertinent to our work, \citet{darian2023enacting} employed \textit{Data Feminism} principles to elucidate how nonprofit organizations utilize advocacy data. Their study, based on semi-structured interviews with 25 staff members from nonprofit organizations, also extended the framework to encompass social good within the realm of the data economy. 

% why use it? what does it enable us to do? how are we using it? and why is it an appropriate framework?
We used the \textit{Data Feminism} framework in our study because of its applicability to AI4SG settings, which emphasize collaborative engagements among multiple stakeholders with the goal of developing data science solutions. Drawing inspiration from the first principle, Examine Power, we discern power relations among AI4SG stakeholders, external influences shaping the objectives of community organizations, and the extent to which these goals are reflected in the realized project outcomes. In tandem, the fifth and seventh principles, Embrace Pluralism and Make Labor Visible, have informed our second research question, focusing on the often-overlooked contributions and labors of community organizations in the discourse around AI4SG. The fourth principle, Rethink Binaries and Hierarchies, has guided our examination of the practical realities of AI4SG projects on the ground, prompting an evaluation of whose priorities are accorded significance. Lastly, we advocate for \textit{data co-liberation} as the overarching objective for AI4SG partnerships, centering the co-leadership of community organizations in data-driven projects. %, positing that such a paradigm shift enhances the outcomes of AI4SG collaborations for the benefit of all stakeholders involved.

%% file: 3-Methodology.tex
\section{Methodology}
Our study aims to understand community partners' perspectives on participation in AI4SG projects and their contributions. Below, we detail our data collection and analysis approach, considering limitations, authors' positionality, and research ethics.

\subsection{Recruitment and Participants}

We collected a list of community organizations involved in AI4SG projects from publicly available websites and reports, focusing on projects that were still ongoing or had already been completed. We then emailed publicly available organizational email IDs for contacts of staff involved in the projects who might be interested in participating in a one-hour Zoom interview. We offered a \$25 Amazon gift card as compensation. Interested participants were asked to complete an online form to provide informed consent and basic information, such as their role in the organization. We reached out to 50 organizations in total and 14 of them responded. With additional snowball sampling \cite{Naderifar_Goli_Ghaljaie_2017}, we had a total of 16 participants and reached data saturation after around 12 interviews~\cite{Pandit_1996}, where further interviews added little new information.

\input{tables/participants}

Table \ref{tab:participants} summarizes our participants' basic information. They worked in various kinds of community organizations, including non-profits, international organizations, and governmental agencies. Some occupied managerial roles like project manager and chief technology officer, while others interfaced closely with the community as field officers, with experience ranging from two to more than 10 years. In total, our participants represented 15 organizations, spanning across social domains such as agriculture and public health. Organizations were headquartered around the world, such as India, Switzerland, and Chile.  %For example, P05 worked with a university lab to develop a Reinforcement Learning algorithm for scheduling health promotion calls in India. P04 worked with a team of data scientists over a summer to develop a machine-learning model that prioritized environmental complaints for a governmental agency in Chile. P12 worked with a technology company to use language models to detect hate speech and misinformation in elections globally (though the organization was based in the U.S.). 

Participants represented 14 different AI4SG projects with various timelines, funding structures, and partnership forming processes. While many of them had worked on or were working on multiple AI4SG projects, we focused on one specific AI4SG project in our interviews to get in-depth insights into participants' experiences. Participants were still encouraged to speak about their experiences with other projects if desired. Two pairs of participants worked on the same AI4SG projects, each occupying different roles (P01 as a field officer and P03 as a domain expert in one project; P08 as a domain expert and P13 as a project manager in another project). The projects all focused on applying AI technologies, including machine learning models, image and video classification, and large-language models. For example, one project built a machine learning algorithm estimating the level of severity and relevance of incoming environmental complaints for government workers. The collaboration was part of a summer program limited to three months. The community organization applied to the program with a project idea in mind, and they were matched with a team of data scientists and graduate students. Another project developed an algorithm predicting disruptions in international trade due to natural disasters like hurricanes. This project was a longer-term collaboration initiated by an external research institute.
%In a project related to agricultural product price estimation in India, P01 worked as a volunteer consultant with agricultural expertise and a technology background, while P03 served as a field officer closely connected to the farmer community. In a project related to human and bee conflict in urban areas in India, P08 worked as an senior ecologist with more than 10 years' experiences and expertise in bee behaviors, while P13 worked as a research fellow in biology playing a supporting role in data collection and analysis. We wanted to capture multiple perspectives within the same projects, but it is not the focus of our study. 

\textbf{Limitations.} One limitation of our recruitment approach and subsequent sample was that we likely ended up focusing on projects that had gotten enough publicity or had a strong online presence, which may correlate with having more resources or more positive outcomes. %This might result in sampling more successful projects and more positive viewpoints about AI4SG. 
Furthermore, since we restricted our search to websites and reports that use English, we did not find projects featured in other languages. 

\subsection{Semi-structured Interviews}

Inspired by ~\citet{charmaz_constructing_2012}’s interviewing techniques, we asked open-ended questions and provided ample space for participants to elaborate on their perspectives without imposing a strict interview procedure. The first author conducted all interviews over Zoom. We used Zoom to generate transcripts automatically, and the first and third authors collaboratively reviewed the transcripts and corrected any errors before the data analysis stage.

During interviews, participants were asked to describe the AI4SG projects they were involved in. Then, the first author probed deeper into the specific aspects of their motivation in participation, problem identification process, solution identification process, engagement with the AI4SG teams, success evaluations, and their opinions on AI4SG as a field. Each interview lasted 55 to 65 minutes and occurred between August and October 2023. 

This study has been approved for exemption by Harvard's IRB office under protocol \# IRB23-1095 and the Brookings Institution. Participants' informed consent was collected in two stages before the interviews: when they registered interest with a Google form and right before we started the interview over Zoom. They were informed of their voluntary involvement in the project, their ability to opt out at any stage, and that they might be anonymously quoted in a research publication. All data presented in this paper has been anonymized, and none of our participants chose to opt out of the study.

\subsection{Analysis}

We used thematic analysis to analyze the transcripts, involving both inductive and deductive components~\cite{braun_using_2006, Braun_Clarke_2019}. Keeping our research questions in mind, which were inspired by the \textit{Data Feminism} framework, the authors collaboratively identified themes through an iterative process. Three authors first independently coded one sample interview and then consolidated the codebook over several meetings to determine the granularity and accuracy of codes. Then, the first author coded all the rest of the transcripts using the same level of granularity. The second and third authors each reviewed half of the transcripts. The first three authors then resolved disagreements and finalized the codes over several meetings. The authors used \textit{NVivo} for coding.

After reviews and discussions, we arrived at eight key themes: "intrinsic motivations for participation," "external factors prompted participation," "participants provided critical contributions for success," "project outcomes fell short of expectations," "impact metrics misaligned with org goals," "orgs updated and diminished goals," "orgs saw diminished goals as valuable," and "partnership aspirations." These categories prompted us to further engage with the framework of \textit{Data Feminism} in order to highlight power dynamics and often-overlooked perspectives in our findings. 

\subsection{Positionality}

%\hongjin{need edits from collaborators}

The interdisciplinary composition of the authors' expertise and experiences shaped our approach to the study. Collectively, the authors have 40+ years of experience studying and building technologies in the social good space, including AI technologies. All authors live and are based in institutions in the US but have conducted research across North America, Europe, Asia, Latin America, and Africa. This shapes the perspectives we come from and are exposed to, but also sensitizes us to differences in contexts across borders. The first author has several years of experience working as an AI4SG practitioner in the domains of international development and environmental sustainability. These experiences help her understand the capacities and limitations of AI technologies and the constraints faced by AI developers. These experiences also motivated the research questions that we asked in this work as she contended with tensions and challenges that arose from working with community organizations.

% Despite diverse personal and professional backgrounds, we also acknowledge our viewpoints heavily bias towards Western values as all authors are based in institutions in the U.S. 

%% file: tables/participants.tex
%Please add the following packages if necessary:
%\usepackage{booktabs, multirow} % for borders and merged ranges
%\usepackage{soul}% for underlines
%\usepackage[table]{xcolor} % for cell colors
%\usepackage{changepage,threeparttable} % for wide tables
%If the table is too wide, replace \begin{table}[!htp]...\end{table} with
%\begin{adjustwidth}{-2.5 cm}{-2.5 cm}\centering\begin{threeparttable}[!htb]...\end{threeparttable}\end{adjustwidth}
\begin{table}[!htp]\centering
\caption{Participant Information}\label{tab:participants}
\resizebox{\textwidth}{!}{
\begin{tabular}{lllllll}\toprule
ID &Location &Org Type &Partners &Role &Experience &Project Domain \\\midrule
P01 &India &Independent &Universities &Domain expert &10+ years &Agriculture \\
P02 &Switzerland &International &Fixed-term Program &Project Manager &10+ years &Connectivity \\
P03 &India &Non-profit &Universities &Field Officer &5 - 10 years &Agriculture \\
P04 &Chile &Governmental &Fixed-term Program &Project Manager &2 - 5 years &Environment \\
P05 &India &Non-profit &Universities &Project Manager &2 - 5 years &Public Health \\
P06 &US &International &Volunteers &Project Manager &10+ years &Development \\
P07 &US &International &Universities &Domain Expert &10+ years &Development \\
P08 &India &International &Universities &Domain Expert &2 - 5 years &Environment \\
P09 &US &International &Industry &Domain Expert &10+ years &Development \\
P10 &Geneva &International &Industry, Universities &Data Science Manager &5 - 10 years &Refugees \\
P11 &Colombia &International &Universities &Project Manager &10+ years &Development \\
P12 &Kenya &Non-Profit &Universities &Domain Expert &5 - 10 years &Environment \\
P13 &India &Non-Profit &Industry, Universities &Project Manager &2 - 5 years &Environment \\
P14 &US &International &Industry, Universities &Project Manager &2 - 5 years &Democracy \\
P15 &New Zealand &Non-Profit &Industry, Universities &CTO &5 - 10 years &Culture \\
P16 &Germany &Non-profit &Universities &Project Manager &2 - 5 years &Public Health \\
\bottomrule
\end{tabular}}
\end{table}

%% file: 4-Findings.tex
\section{Findings}

\begin{figure}
    \centering
    \includegraphics[width=0.95\linewidth]{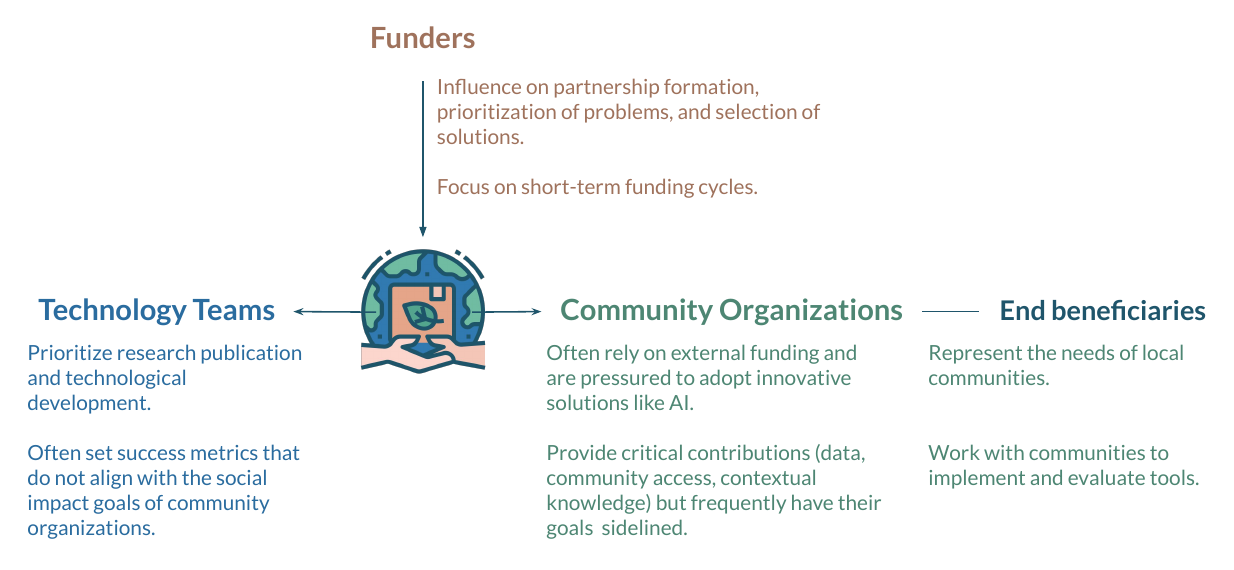}
    \caption{Relationships among stakeholders in AI4SG partnerships. Icon made by GOWI from www.flaticon.com.}
    \label{fig:stakeholders}
\end{figure}

In what follows, we utilize the \textit{Data Feminism} framework to structure our findings. Investigating the first research question on community organization members' motivations to participate in AI4SG projects and the factors that influence their motivations, we examine and reveal power dynamics among AI4SG stakeholders (Figure ~\ref{fig:stakeholders}). We then detail the critical (yet often overlooked) infrastructural and intellectual contributions that community organization members provide to the partnerships, addressing the second research question. We conclude with a depiction of the realities of AI4SG initiatives on the ground and community organization members' perspectives on their outcomes over time, focusing on the third research question.

\label{findings}

\input{4-1-motivation}

\input{4-2-contribution}

\input{4-3-evaluation}

%% file: 4-1-motivation.tex
\subsection{Power Dynamics and Factors Shaping Participation in AI4SG Projects}
\label{motivations}

% - funders - orgs depend on external funding, shape project formulation and solutions
% - social beliefs - interested in innovation and new tech
% NEW SUBSECTION overarching funding and hype factor
% - funding
% - hype
% Breakdown of individual and organizational motivations
% initiators - specific tech needs, want tangible products and learning
% along the riders - put into projects, need to do it for org, expect to reduce their own workload
% strategists - approached by tech teams, want to maintain partnerships, boost funding, access to computing, expect to generate new ideas and publicity

\begin{quote}
    \textit{``Data Feminism begins by analyzing how power operates in the world today.''} --- \textit{Data Feminism}~\cite[p.47]{dignazio_data_2020}
\end{quote}

The first principle of \textit{Data Feminism}---Examine Power---calls for a close look into data science by whom, for whom, and with whose interests and goals in mind. In this section, we apply this principle to analyze community organization members' motivations and goals, revealing the power dynamics among stakeholders in AI4SG projects, including the community organization members, technology teams, funders, and the intended end beneficiaries. Overall, we found that community organizations' participation in AI4SG projects was heavily influenced by external factors, especially funding agendas and broader beliefs about the potential of AI technologies. Offering a breakdown of community organization members' specific motivations and goals, we further show how internal power relationships within the organizations influence individuals' participation and perspectives in AI4SG projects. % Internal organizational factors also shaped individual participants' involvement and expectations in AI4SG partnerships, depending on their specific role.

\subsubsection{Influence of funding and belief in the promise of AI on AI4SG collaborations}
% This means that funders have significant sway over organizations in AI4SG project formation and solution identification. Funders hold a lot of power in determining the problems and solutions of AI4SG partnerships. Since funders are a few degrees away from the day-to-day adoption of AI technologies, they might be more optimistic about AI technologies and expect more from the projects than community organizations and technology teams themselves. 

The community organizations in our study share the characteristics of a mission-driven organization: limited resources, grant-driven financial models, impact measurement requirements, and reliance on external technical expertise~\cite{bopp_disempowered_2017, erete_storytelling_2016, Kumar2018towards}. These factors contributed to \textbf{the pervasive influence of funders} on community organizations' involvement in AI4SG. Our interviews revealed various funding sources, including government agencies (P15) and private companies offering grants for AI4SG projects specifically (P06, P09). They determined, to a large extent, how partnerships were formed, what problems were prioritized, what solutions were explored, and how long the project lasted (cf.~\cite{saha2022commissioning,bopp_disempowered_2017,darian2023enacting, Sum_2023}). For example, P09's project involved building a language model to analyze policy documents. Together with colleagues, they applied for a grant provided by a private technology company and submitted multiple ideas of problems that could be solved by AI-related technologies. %While P09's organization was driven by a desire to explore AI solutions (hence the application to the AI-specific grant), 
The technology company made the final decisions on which ideas to pursue, even though they might not align with the community organization's priorities: 

\begin{quote}
    \textit{``For [our organization], the other idea was a priority. But in the context of the interaction with the [industry partner] on how to pick an idea, it was not prioritized in that way. We had a number of ideas, and then it was the [industry partner] that sort of decided which ones they would pick.'' (P09)}
\end{quote}

Funding for AI4SG projects in our study was also often fixed within a short time frame, impacting what the partnerships could realistically achieve and whom they benefit. In P09's case, the grant provided three months of technological support from a student research team. Because the grant only lasted for three months, they had to constrain their ideas to ones feasible within that time frame. The project resulted in a publication on the model the technology team built, but the model was never implemented because the funding ended after three months. In another example, P08's project aimed to track bee activities in urban areas in India using computer vision techniques. The project received a year's funding, but it was not enough time to develop a workable model because the particular task was so difficult. P08 emphasized that projects aiming to address complex social issues \textit{``cannot be done in such a short duration.''} In this case, short-term funding did not benefit community organizations as much as it benefited researchers and students who were able to publish about their models and research processes. 

Funding also often constrained the product of collaborations to take the form of what funders had in mind. As P09 acknowledged, while \textit{``there are many ways that you can solve the problem, [...] it was all about what [the grant] was offering us at the time.''} Some participants explicitly pointed out how funders' needs determined the project's end product. For example, P06's project on building a humanitarian aid data platform was funded by an industry organization:
\begin{quote}
    \textit{``There's this perverse incentive of responding to the donor. There might be someone at a foundation who loves the idea of this platform and really wants to see it built. [...] And we end up building this platform for the funder with all the things the funder wants to see in it.'' (P06)}
\end{quote}
Importantly, P06 then stressed that the pressure to meet funders' visions could result in a solution that \textit{``never works in the field because no one else wants it in that way.''} %Responding to funders' agenda did not necessarily result in the real-world positive impact that AI4SG projects strived to achieve. 

Another overarching factor affecting the nature of AI4SG collaborations is \textbf{the belief in the promise of AI for social applications}. For example, P01 worked with an academic research team to build a machine learning model predicting crop market prices for farmer collectives in India. Reflecting on the limitations of the machine learning model, they believed that AI4SG was in \textit{``this hype phase''} and \textit{``it is really difficult to separate reality from people's wishes.''} This hype factor could influence community organizations both directly and indirectly. In the former case, many participants pointed out an internal organizational incentive to adopt AI and other innovative solutions to assist with impactful but time-consuming work. For example, P11's project on building an AI chatbot serving farmers was a direct response to their organization's \textit{``constant reach for what is new out there.''} In the latter case, this hype factor affected funders' beliefs in the potential of AI and funding agendas, which then shaped AI4SG partnership priorities as described above. 

% \textbf{Social Beliefs.} Lastly, all stakeholders are influenced by the social beliefs about AI and technology innovation more broadly, and its promised capacities for solving social issues. Many of our participants pointed out an organizational tendency to adopt innovative solutions. For example, P11 pointed out that ``we are always on the reach of what is new out there that we can tailor to the work.'' In P09's organization, exploration of new technology for development is ``part of [our organization]'s DNA, and there's always been a strong investment in innovation, be it AI or other technologies.'' 

% While acknowledging a need to innovate, our participants also pointed out the hype around AI4SG. P01 shared that ``it is still this hype phase where I think it's really difficult to separate reality from people's wishes.'' P14 agreed and shared their opinions on AI adoption in their organization:

% \begin{quote}
%     \textit{“It is very blurry. Everyone talks about it, but there's very few people who actually get concrete actionable working outcomes with that yeah you can say that the word 'good' was used most times than 'bad'. Often I think the most important is to have AI in the title and no one really cares what it does down the line and you just have a report of 300 pages. And it's useful for no one.” (P14)}
% \end{quote}

% These larger social beliefs at play influence community organizations directly as well as indirectly through other power relations within the AI4SG stakeholder map.

\subsubsection{Community organization members' motivations and goals}

Centering members of community organizations in this wider context of constrained funding and the belief in the promise of AI, we found a diverse range of motivations that prompted their participation in AI4SG. We broadly group our participants into three categories--the Initiators, the Implementers, and the Strategists. Table~\ref{tab:groups} summarizes the three categories and their characteristics. 

\input{tables/groups}

\textbf{Group 1: The Initiators}
\label{group1} 

In the first group, participants initiated the projects on behalf of their organizations. They either applied for fixed-term funding initiatives (P06, P11, P16), summer programs (P02, P04), or directly hired technology contractors or companies (P12, P14) to fulfill their technology needs. They often occupied managerial or analyst positions and could directly influence AI adoption and usage decisions. We highlight two main motivations for their involvement in AI4SG projects. 

First, they had \textbf{a specific problem (and sometimes a specific solution) in mind, but needed access to technology capacity not available internally}. In P14's case, they had a clear goal in mind: to track misinformation and detect hate speech online with machine learning, and they wanted deployable software that could be used by their end beneficiaries. After surveying different partnership options, they hired a technology company to build the software. They thought hiring a company would enable them to customize the software with maximum flexibility and autonomy. In this partnership model, the expected outcomes were commonly \textbf{tangible products in deployment}. For example, P04 \textit{``wanted to have something in production.''}  P06 initiated their project \textit{``with the goal of eventually publishing a global public software.''} P12 \textit{``expected at least some MVP [Minimum Viable Product] of sorts''} and that the team could \textit{``actually build something; otherwise, what’s the point?''} 

Second, many valued the \textbf{exploration of innovations and interdisciplinary collaborations} for the long-term future of their organization and addressing social issues, often expressing a desire to catch up with the private sector. This motivation was influenced by a larger organizational incentive to adopt innovations (cf.~\cite{Saha2022towards}). As P06 stressed, \textit{``there is increasing skill and ability to use data within humanitarian work in general, but we still lag behind the private sector, and that causes more reason for collaboration.''} Similarly, P14  \textit{``saw a lot of tools developed by the private sector that would make a great impact''} in their own domain. %Since they saw their participation in collaborations as a long-term investment in their domain, they approached the technology teams with a high level of flexibility and reliance on the technology teams to identify and develop solutions. %without necessarily expecting a tangible software product. %For example, \added{P16's project had an ambitious goal of creating a computer vision-based monitor for child malnutrition in developing countries,} and they relied on their partner from an academic lab to \textit{``bring the novel solutions that we have not considered before.''} 
Many participants highlighted the value of the \textbf{learning process} in pursuing innovative solutions, even if they might not be implemented in real life. For example, P02 initiated a project to map regions of the world without internet connectivity and applied to work with a summer program. The project resulted in a presentation on several country-specific machine learning models. While these models were not deployed, P02 emphasized the importance of \textit{``continuous learning''} and \textit{``[engaging] in things that are new, that are slightly outside our area but still connected to that.''} % In the process of pursuing AI solutions, P04 “learned a bit more” about new technologies and “putting in production this kind of models” and found it “personally very rewarding.” 

\textbf{Group 2: The Implementers}
\label{group2} 

In the second group, the interviewees themselves were not involved in the partnership formulation stage, but they were brought into the projects as a part of their existing responsibilities. They often occupied the position of field officers or domain experts, and they mediated between the end beneficiaries, the technology team, and the management team of their organizations. They also performed translation tasks to address technological barriers in local communities (cf.~\cite{Sum_2023}). 

While they might not have the agency to decide what problems to work on, several participants expressed a general motivation to \textbf{reduce existing their own manual workload} by participating in AI4SG projects. For example, P13 worked with a university team to develop an algorithm to classify bee dancing patterns, as an ecology expert in the project. They were excited by the potential of AI to automate \textit{``most of the tedious tasks of watching hours of [bee dancing] videos.''} 

%Participants also expressed a personal joy in learning about the new technologies. For example, P05 shared that the project was “an eye-opening thing because I have never been in a role where I was managing technology plus community.” 

Given the internal power dynamics within community organizations, the views of individual staff members may not represent the overall motivations and goals of the organizations (cf.~\cite{maxwell_data_2016, kawakami_studying_2023, Saha2022towards}). We also found that while some participants in this group expected the projects to be a learning process, their organizations sometimes expected more concrete outcomes. In P09's words:

\begin{quote}
    \textit{``For me personally, it was an experiment. So it was mostly about having a proof of concept, working with [the technology team] to show that this type of approach can help address the issues that we did have. So my expectations were just at that level. Within the organization, there were, I think, higher expectations that we would indeed have a turnkey solution. We would have something where colleagues could upload the documents, press a button, and it gives them the spreadsheets.'' (P09) }
\end{quote}

%Not unique to participants from Group 2, this tension among different stakeholders within the same organization is also present in Group 1 participants who initiated the projects themselves on behalf of the organization. For example, P02 described the need to bypass supervision and lengthy organizational procedures to apply for the AI4SG program in time, because the project “really comes on top of everything else that we have on our desk” and it was hard to have “the project to be recognized as something valuable and worth investing as an organization.” 

\textbf{Group 3: The Strategists}
\label{group3} 

In the third group, neither the participants nor their organizations initiated the projects, but they decided to participate when approached by external technology teams. For example, in P10's case, an academic research team \textit{``stumble[ed] upon our datasets on our platform''} and approached P10 to form a partnership using the datasets to answer a research question related to immigration patterns. While the name \textit{``strategists''} implies a high level of agency (which is true to some degree---participants got to choose to participate or not), the external factors influencing their decisions to participate were outside of their control. We highlight a couple of these factors below.

%  % Participants from this group were also individuals with decision-making power in their organizations, occupying roles such as the Chief Technology Officer (CTO) or manager.
% These projects might not be a top priority for the organizations, but external factors prompt them to start a new partnership or maintain existing ones.

%First, participants see value in building long-term relationships with the “actual human beings” of the institutions that they work with, since “it's better for us to work with people that we get along with” (P15). P15 also emphasized the value of “reciprocity:”

% \begin{quote}
%     \textit{“It works both ways. Right? Like people. We've gone to people and asked for their support, and they've supported us, and now people are coming to us and asking us for our support. And there's that reciprocity, and it's important that we do that.” (P15)}
% \end{quote}

First, participants saw value in building and maintaining long-term relationships with external technology teams. These relationships were valuable to them for \textbf{access to a network of interdisciplinary stakeholders and for generating new ideas and future projects}. Underlying this motivation was a belief in the promise of AI technologies and an organizational incentive to innovate, echoing the sentiment of the Initiators.   
%In P08’s case, while they had an initial project idea related to land use effects on bee colony aggregations, they ``learned from [the technology team]'' and “jointly discussed” a new idea about the automated video detection of bee dancing. P08 shared that the idea “came in through the interaction” with their collaborator, and “if it was somebody else, I don’t think we would have tried to do that.” 

Second, participants were sometimes incentivized to \textbf{partner with prestigious institutions to boost their own credibility for funding opportunities}. Our participants' organizations often needed to go through frequent funding applications (cf.~\cite{bopp_disempowered_2017}), and were under scrutiny for credibility regarding technology innovations (cf.~\cite{erete_method_2023}. As P15 put it, \textit{``having someone, a well-respected scientist from like Oxford or Cambridge, just name whatever is popular, definitely helps raise the profile.''}

% Other motivations for maintaining AI4SG partnerships included \textbf{gaining or maintaining access to computing infrastructure,} especially when the solutions require a large amount of computing resources. With most computing infrastructures, like cloud storage and GPUs, owned by dominant technology companies, social sector organizations found themselves reliant on these resources for innovations. %Since these resources can be costly to acquire and maintain, social sector organizations --- often resource-constrained --- might choose to participate in a project just to gain access to or maintain these resources. 
% For example, in P10’s case, they had wanted to build an AI-based application that required \textit{``a big, big server.''} When a technology company with sufficient computing power offered to collaborate, P10 agreed to participate, driven by the need for computing infrastructure to realize their vision. 

%% file: tables/groups.tex
%Please add the following packages if necessary:
%\usepackage{booktabs, multirow} % for borders and merged ranges
%\usepackage{soul}% for underlines
%\usepackage[table]{xcolor} % for cell colors
%\usepackage{changepage,threeparttable} % for wide tables
%If the table is too wide, replace \begin{table}[!htp]...\end{table} with
%\begin{adjustwidth}{-2.5 cm}{-2.5 cm}\centering\begin{threeparttable}[!htb]...\end{threeparttable}\end{adjustwidth}
\begin{table}[!htp]\centering
\caption{Breakdown of community organization members' motivations and goals}\label{tab:groups}
\resizebox{\textwidth}{!}{
\begin{tabular}{llllll}\toprule
Group &Participants &Characteristics &Key Motivations &Key Expectations \\\midrule
Group 1 &P02, P04, P06, P11, &The participants initiated the projects &Specific technology needs &Tangible products \\
The Initiators &P12, P14, and P16 &on their organizations' behalf. &Explore innovations &Learning experiences \\
& & & & \\
Group 2 &P03, P05, &The participants were brought into &Fulfilling existing &Reduce manual \\
The Implementers &P09, and P13 &the projects by their organizations. &responsibilities &workload \\
& & & & \\
Group 3 &P01, P07, P08, &The participants were approached &Maintain long-term partnerships &Generate new ideas \\
The Strategists &P10, and P15 &by external technology teams. &Boost funding credibility &Publicity \\
% & & &Access to computing & \\
% & & &infrastructure & \\
\bottomrule
\end{tabular}}
\end{table}

%% file: 4-2-contribution.tex
\subsection{Community Organization Members' Contributions and Labor}
\label{contributions}

\begin{quote}
    \textit{``Embracing pluralism in data science means valuing many perspectives and voices and doing so at all stages of the process—from collection to cleaning to analysis to communication.''} --- \textit{Data Feminism}~\cite[p.130]{dignazio_data_2020}
\end{quote}

We apply \textit{Data Feminism}'s fifth and seventh principles---Embrace Pluralism and Make Labor Visible---to highlight and honor the (often overlooked) contributions and labor of community partners. We found that participants provided critical data and access to the local communities--two necessary conditions for project success. They also contributed intellectual efforts instrumental in developing AI technologies, providing feedback on model development, system maintenance, and research publications for technology teams. % Last but not least, our participants had a clear sense of what might be successful on the ground as well as the ethical considerations and responsible practices of AI solutions. % If provided the chance to lead or co-lead the AI4SG projects, our participants would have been able to improve the project outcomes. 

\subsubsection{Providing the necessary conditions for success}

First, our \textbf{participants provided access to valuable data that was otherwise inaccessible.} AI technology relies on data~\cite{wang2022whose}, and data collection often entails extensive effort over a long period of time (e.g.~\cite{sambasivan_deskilling_2022, miceli2022data, miceli2020between}). For example, P12’s project aimed to estimate grassland quality using remote sensing data. To provide training data for the model, they \textit{``actually had to go measure grass […] and that was like a thousand data points.''} We additionally found that collecting such data had overhead beyond the collection itself. P02, for instance, shared that data collection could be \textit{``tricky''} because they needed to collect administrative data from multiple countries, each requiring a separate approval process.

Since many projects involved vulnerable populations like refugees, our participants emphasized the importance of data privacy and the special handling of sensitive data so that they could be used responsibly. Community organization members were diligent about meeting reporting requirements, which were a form of data work that could be invisible to their technology partners. For example, P16’s project aimed to estimate child malnutrition from images, and they provided \textit{``data [that] not everyone can have''}:

\begin{quote}
    \textit{``To collect data, you need the ethical approval from the government, from the health industry, we have done that. We've gone through the pain. It is basically expensive to collect.'' (P16)}
\end{quote}

Besides data collection, the data-sharing process entailed data cleaning and engineering work. For example, P07 wrote data queries and shared code that cleans the data, so \textit{``[the technology developers] don’t need to deal with this.''} P11 also mentioned they \textit{``would work on all data crunching and data mining during the week, and it is a lot of work that goes behind that.''}

Second, our \textbf{participants provided access to community platforms where technology teams could implement their ideas.} In a project on Indigenous language preservation, P15's organization worked closely with the local Indigenous communities and was able to provide \textit{``the way in''} for their university partners. As P15 emphasized, 

\begin{quote}
    \textit{``I think the key is what we bring to the relationship, and what we bring is a close connection to the communities. It was essentially the holders of the data that these researchers wanted to study. And they don't have that. No academic institution has links back to the community like a community organization.'' (P15)}
\end{quote}

Working with local communities to implement tools took significant effort. %Community organizations must build trust in their community and maintain meaningful relationships~\cite{Sum_2023, Harrington_2019, LeDantec_Fox_2015, erete_method_2023}. Playing the critical role of a broker~\cite{hou2017hacking, alvarado2023mobilizing} and technical translator~\cite{Sum_2023}, 
Participants worked to build trust in their communities, maintain meaningful connections even amidst the introduction of technology, obtain buy-in from community members (P03, P05), explain what the projects were about (P05), convert research findings into local implementation and field tests (P14), get feedback from community members about the projects, and provide operational support on the ground (P01, P03). For example, when testing out a phone-based AI system to provide maternal health information to mothers, P05 shared:  

\begin{quote}
    \textit{“There's a lot of apprehension among people to give phone numbers to subscribe to you, because first of all, there were [scams] in these health programs. But then we had to talk to people, get their trust in the program. […] That's an extra effort that we have to do. We thought that if we just roll [the AI solution] out people might respond and all, but then that's not the case. People don't respond.” (P05)}
\end{quote}
In this case, P05's \textit{``extra effort''} in the local community was crucial for the local implementation of the project. %For any AI4SG projects to claim ``real-world'' impact, they need access to the real world, which the community organization members facilitate. %; they serve as a bridge between the local community and the technology teams and do the work to make interventions work. 

\subsubsection{Intellectual contributions to AI development}

Our participants also provided both technical and non-technical guidance throughout the projects. These intellectual contributions were crucial for ensuring practical solutions in real-world situations. We highlight a few of them below. 

First, our participants provided \textbf{domain expertise and local knowledge that guided the high-level direction of AI development}. For example, in a project related to building a bee activity map, P08 pointed out that \textit{``[the technology team] is not familiar with this landscape so we had to clarify a lot of things''} like \textit{``checking whether a forest is a forest, etc.”} P02 also emphasized that they \textit{``always remind [the technology team] that we need to keep that perspective of what is the product that we want to deliver in the end and what is the purpose that product is going to serve.''} In providing this high-level directional guidance, our participants made sure that the technology team, who could otherwise often get \textit{``too much into the specifications in the technology''} (P02), made model choices that \textit{``are most relevant.''}  

Second, our participants also provided \textbf{technical model evaluation and feedback}. Several participants in Group 1---the Initiators (Section~\ref{group1})---in particular had a technical training background themselves and gave specific guidance on model development. For example, when working with a summer student team to develop a machine learning algorithm for prioritizing environmental complaints, P04 shared:

\begin{quote}
    \textit{``I always point out how the training was doing because the first time we met, they brought very good metrics. I don't know, like, an accuracy rate of more than 90\%. So I was very like, that sounds weird. Are we over-fitting? I feel like we are over-fitting. Like, because the guys, again, they were brilliant really, but they never had experiences in real-world machine learning modeling.'' (P04)}
\end{quote}

%This quote also demonstrates the limitations of some technical teams consisting of relatively inexperienced university graduates. 

Besides high-level strategic guidance and technical evaluation, our participants made other intellectual contributions. For example, P09 and P13 played an active role in supporting research publications and outputs. Furthermore, we found that all participants committed to regular and frequent online and offline communication across interdisciplinary teams, with some often contributing to project management tasks (P10, P11, P12, P16), convening the right teams and stakeholders (P07), and writing funding applications (P06, P15)---all of which require \textit{``lots of thinking''} and \textit{``lots and lots of time''} (P02).

%If they were given the chance to share their knowledge early on in the process and throughout the project, the chance of success would be a lot higher. 

%%%%% other aspirations

% \textbf{Last by not least}, our participants advocate for inclusive AI and “technology for everyone by everyone” (P09): 

% \begin{quote}
%     \textit{“It's good to see the interest from the private sector companies. But I hope that it's just not a way of, you know, selling their products and gaining market share. I mean, we've seen that before. My inspiration is that we actually make this amazing new technology work for people who would need it the most, not the people who really have everything and want more. Make technology work for everyone and by everyone.” (P09)}
% \end{quote}

% P05 shared this sentiment and thought that:

% \begin{quote}
%     \textit{“it is sort of a responsibility that everyone who deploys a certain technology to go that extra mile and do a sort of tutoring of people, and telling them what [the AI solutions are] really doing and how they can probably make the best use of the technology that's been given to them.” (P05)}
% \end{quote}

%% file: 4-3-evaluation.tex
\subsection{Shifting Expectations of AI4SG on the Ground}
\label{evaluation}

\begin{quote}
    \textit{``What gets counted counts.''} --- Joni Seager cited in \textit{Data Feminism}~\cite[p.97]{dignazio_data_2020}
\end{quote}

\textit{Data Feminism}'s fourth principle calls for challenging binaries and hierarchies embedded in counting and classification systems. What gets counted and measured shapes the priorities of AI4SG projects and determines who benefits from them. In this section, we investigate what success metrics got prioritized and whether the outcomes of AI4SG projects met the community organizations members' goals. We found that AI4SG projects often fell short of community organizations' expectations of deployment, and the success metrics prioritized by technology teams did not help further community organizations' goals. Despite initial disappointment, our participants still saw diminished goals as valuable and recognized the competency of technology teams. Lastly, we share participants' aspirations for collaborations that would help avoid disappointment in the future. 

\subsubsection{Realities of AI4SG falling short of expectations}

While many participants expressed optimism about the \textit{potential} of AI solutions, \textbf{the actual outcomes of their projects often failed to meet their initial expectations}. The most prominent case was when the organizations initiated the projects for a specific need and hoped for a \textit{``game-changer''} solution (the Initiators, Section~\ref{group1}), but there was no real-world deployment due to limiting factors like funding constraints. For example, P12 partnered with an industry team to develop software for grassland quality estimation using remote sensing data. However, they did not see the project come to deployment since the technical support stopped after the funding ended after a year. Reflecting on the project, P12 shared that while the partnership was \textit{``great for bouncing ideas back and forth,''} it was \textit{``insufficient for actually building anything of real substance.''}

Besides funding constraints, incentive misalignment between community organizations and academic or industry partners also contributed to community organizations' goals not being prioritized (cf.~\cite{Kumar2018towards}). For example, P07's organization partnered with universities and emphasized that \textit{``[research labs] don’t deploy the solutions; they finish the paper, and that’s it, because their goal is to research, so when research is done, most probably there is no continuity''} (P07). This motivation to publish meant that projects were commonly evaluated by model performance or other standard metrics for AI publications. While strong machine learning model performance led to publications for AI teams and bolstered funding applications for community organizations, model accuracy did not translate to other outcomes that community organizations also cared about. As P15 emphasized, when working with a university partner on a language model specific to Indigenous languages, publication was \textit{``not a priority for us.''} They cared more about \textit{``what we are doing for the Indigenous community''} in terms of \textit{``how many [of them] are actually using the model that we've built.''}. However, in P15's case, the partnership led to publications rather than the Indigenous community adopting the model.

When working with private industry partners, our participants expressed frustration about the tension between industry partners' profit incentives and their own aspiration for public interest. For example, when developing a model for predicting immigration patterns with an industry partner, P10 shared that \textit{``it's very, very difficult to work with private sector partners, because they have the interest of selling the software to us, and we have the interest of tweaking whatever they have for serving our needs.''} In addition, working with the private sector often entailed \textit{``legal conundrums''} that prevented the social sector organizations from actually purchasing and deploying their products when they intended to (P09, P10). 

Reflecting resource constraints and the misalignment in incentives, \textbf{only two out of the 14 projects resulted in real-world deployment} (P11, P14). Among organizations that did reach deployment, participants still saw project success not as model accuracy, but as what the tools enabled them to do concretely. For example, P11's project resulted in an AI chatbot that answered requests for information from farmers and reduced the staff's manual workload. P14 worked with a software company to customize a platform for government agencies to detect misinformation during elections. Again, metrics like model accuracy and publications were not central to their conceptualization of impact. 

Among participants who did not expect deployment, they were still disappointed when the project duration was too short to solve their problems (P08), or there was a lack of engineering support and training for the organizations' staff who would interface with the tool (P12). 

% For example, P08 emphasized that “often a problem like this cannot be done in such a short duration,” and ``we would have liked some more follow-up meetings which didn't happen.'' These cases of disappointment reflected the lack of power community organizations held in determining AI4SG priorities. 

\subsubsection{Seeing diminished goals as valuable}

Despite initial disappointments, our participants appeared forgiving and dynamically updated their expectations over the course of the projects. The updated expectations were often diminished from the initial ones, such as changing from expecting a tangible product to appreciating the learning process. In P06’s case, their initial stated goal was to build a humanitarian aid data-sharing platform. They shared how their expectations changed as the project progressed slower than they expected: \textit{``at that time, I would have thought that by now we'd have a finished working platform. But it always takes longer than you think. And you always learn a lot along the way.''} 

Other than emphasizing the learning process, our participants saw other byproducts from the partnerships as valuable. In some cases, these byproducts were directly useful for addressing the social problems at hand. For example, the process of developing the AI technologies helped our participants further clarify the problem space. P04 pointed out that their short-term summer project \textit{``gave us a better understanding of this big problem of the institution.''} Other participants also acknowledged that learnings from these projects could be translated into other projects. 

In other cases, our participants valued the partnerships for other benefits not directly related to the problems at hand. For example, P12 pointed out that \textit{``publicity''} of the projects could help boost their funding applications. P02 emphasized a \textit{``wow effect''} from their organization's leadership when presenting high-performing models even though they were never deployed. 

% Despite diminished goals and initial disappointment, these cases reflected the organizational and funding factors that sustained our participants' involvement in the AI4SG projects. They often continued to believe in the \textit{potential} of AI4SG projects despite a lack of concrete evidence that they fulfilled the social impact they strove for. For example, in P08's project, they shared that their \textit{``AI counterpart didn't finish their part of the work,''} mostly due to the short funding time frame and the problem's difficulty. However, they still believed that \textit{``the potential [of the project] is huge.''} 

% \hongjin{Catherine also says that technical teams/research teams are not doing a good job of setting reasonable expectations. In haste to perform ``social good'' and publish about how socially good they are, perhaps technical teams are over promising and underdelivering. these other values are valuable and tangible too. be honest and humble about the capacities of AI technologies. as other outputs outside of this paper: pragmatic guides for different stakeholders?}

\subsubsection{Recognizing technology teams' technical expertise}

Despite the crucial intellectual contributions that community organization members provided for the partnerships, our participants continued to emphasize the technology teams' technical expertise in various ways--- both because it genuinely contributed new skills, but also because they felt obligated to in order to obtain funding.

Incentivized to explore and adopt AI solutions, in tandem with limited internal technical capacity, our participants had to rely on external technology teams (cf.~\cite{bopp_disempowered_2017}). Participants saw the technology teams' roles as instrumental because \textit{``they were really doing all the work that we couldn't do.''} (P02) The emphasis on technology teams' expertise was most prominent in the first and second groups of participants---the Initiators and Implementers (Sections~\ref{group1} and~\ref{group2}), where the participants had a specific technology need. %Participants in the third group---the Strategists (Section~\ref{group3}), who did not initiate the partnerships nor had a specific technical need, also demonstrated a tendency to highlight the technical teams’ role in determining success. For example, while P08's project did not meet their expectations, they thought having \textit{``more support from experts from [the technology company]''} would have helped accomplish their goals. %We hypothesize that external factors at play, including fascination with AI technologies, influenced our participants' perceptions of expertise.

Our participants' perception of technology teams' ability to secure funding also influenced how likely they were to claim ownership and leadership over AI4SG projects. In applying for a grant focusing on AI technologies together with a university team, P15 shared that because the university team had a higher chance of securing funding historically, they questioned their own leadership in the project:  

\begin{quote}
    \textit{``It builds into your psyche. [...] I had to go through my psyche of should our organization lead this fund or should we put [our university partner] as a leader, because that would give us a better chance; looking at the data and history, it's usually them who get the funding.'' (P15)}
\end{quote}
This self-questioning impeded P15's decision to put their organization's name down as the leader in the funding application and the project as a whole. In subsequent funding cycles, they were even \textit{``forced to collaborate''} with another organization to access funding, due to the funding agency's perception that the collaboration would improve project efficiencies (P15). However, the collaboration eventually failed due to interest misalignment and the other organization's misunderstanding of the local contexts in which P15's organization was situated.

\subsubsection{``Come to us first'': Aspirations towards improved partnerships}

Our participants speculated that following their aspirations, described below, would help ensure ethical and successful AI4SG partnerships. It is important to point out that all of these factors were present in the two projects that resulted in successful deployment (P11, P14).

First, our participants preferred a \textbf{relationship-first and need-driven approach} to a partnership where they would be brought in as co-thinkers in the ideation stage. Reflecting on various AI4SG projects with different technology teams, P07 shared: 

\begin{quote}
    \textit{``So our dream is that before a research institute decides to do something, they come to us first and ask ‘what do you need?’ Rather than ‘oh I need to use this tool so I need to work with you.’ So hopefully in the future when we work with a research institute, they come to us first and expect, okay, so we are going to research something. But let’s decide together what it is.'' (P07)}
\end{quote}
In both P11's and P14's cases, they had a tangible technology need and played an active and collaborative role in determining the problems and solutions with their technology teams. They also initiated the projects instead of being approached by technology teams. 

Second, participants emphasized the importance of \textbf{leadership of community organizations} in AI-driven projects--as P15 put it: \textit{``it's important that we are the leaders of that---that we're in charge of how those technologies are developed and used.''} Our participants often had a clear sense of what might be successful and whether AI would be the right solution. For example, P07 saw AI as \textit{``just part of the methods that we use''} and emphasized that \textit{``AI will be used if it's fit for purpose.''} If AI is used only because it fits the funders' agenda or an organization's desire for innovation without concrete evidence that it works, AI4SG projects risk overpromising what they can achieve. P14 emphasized the issue of just \textit{``having AI in the title''} without considering its downstream impact: 
\begin{quote}
    \textit{``Often I think the most important thing is to have AI in the title and no one really cares what it does down the line, and you just have a report of 300 pages. And it's useful for no one.'' (P14)}
\end{quote}
Because of P14's active involvement in the solution ideation stage, they were able to deploy their solution and avoid just having \textit{``AI in the title.''} The solution they deployed was a \textit{``simple''} linear regression model, instead of a \textit{``very high-end AI tool that needs the support of super trained analysts.''}

Community organizations' leadership could also help ensure the ethical development of AI technologies for social applications. AI4SG projects often concern sensitive social issues that impact real human beings. Our participants worked closely with local communities they served and could provide important contextual knowledge to mitigate harm to the intended end beneficiaries. P10 made a strong case for community organizations' leadership in problem formation and solution development for ethical reasons: 

\begin{quote}
    \textit{“What I hate the most, and I've been working here for eight years, is when a company comes and says ‘We would like to apply our product into refugees.’ And I was like, they're not guinea pigs. They're human beings with dignity. No, no, don't contact me until I contact you. It's unethical, you know, imposing a product or service or technology. They're not allowed for experimentation.” (P10)}
\end{quote}

In P11's case, they were mindful of the potential harm of promising a solution to people in dire situations without actually delivering help, such as in disaster relief:

\begin{quote}
    \textit{``It's a problem because we're creating an expectation that we're gonna save them and we don't know if we can. So if you set up a system that creates an expectation, you have to do something. Otherwise, you can do more harm than good.'' (P11)}
\end{quote}

% Considering contexts, as \citet{dignazio_data_2020} pointed out, is essential for developing accurate and ethical products. The leadership of community organizations is crucial for ensuring the contexts are considered in AI4SG projects.

Last but not least, our participants also emphasized \textbf{the importance of internal capacity building} for the sustainability of the solution after the AI4SG projects end (cf.~\cite{Winschiers-Theophilus_2010, Unertl_2016}). For example, as P11 demanded, \textit{``we wanted people to come to sit with us for meetings to give us training,''} because \textit{``when we do not incorporate a capacity strengthening element, the solution dies, because we are not able to maintain it, so we are not able to sustain it over time.''} Because of their insistence on requesting training for staff, P11's organization was able to maintain and operate the solution themselves after their university partner left the project. Beyond project success, P05 also stressed the training of staff and local communities as a \textit{``responsibility''} of the technology teams:
\begin{quote}
    \textit{``It is sort of a responsibility that everyone who deploys a certain technology to go that extra mile and do a sort of tutoring of people, and telling them what [the AI solutions are] really doing and how they can probably make the best use of the technology that's been given to them.'' (P05)}
\end{quote}

%% file: 5-Discussion.tex
\section{Discussion}
\label{discussion}

Community organization members' experiences and perspectives in AI4SG partnerships challenge the prevalent narrative promising tangible AI solutions to complex social issues (e.g.~\cite{tomasev_ai_2020}). Utilizing the \textit{Data Feminism} lens by \citet{dignazio_data_2020}, our study uncovers power imbalances among community organizations, technology teams, and funding agendas. Our findings contribute nuanced insights into \textit{where} power imbalances manifest in AI4SG collaborations over time, influencing 1) community organizations' motivations and project expectations, 2) project priorities, 3) project evaluations, and 4) perceptions of technical expertise. These power dynamics, coupled with the enchantment of AI technologies, contributed to projects falling short of their deployment expectations. Despite community organization members finding value in diminished goals and extracting other benefits from partnerships, they shared aspirations for factors that could improve the impact of future partnerships: 1) adopting a relationship-first and need-driven approach, 2) ensuring that community organizations can co-lead early on, and 3) investing in technology capacity building for community organization staff.

Drawing on our findings, we first examine the wider social conditions that generate interest in AI4SG and discuss their implications, calling for a shift of focus from the tool (AI) to the social issues at hand. Following this, we discuss power imbalances in AI4SG partnerships and bring the idea of \textit{data co-liberation}~\cite{dignazio_data_2020} to bear on creating more equitable collaborations. Overall, we argue that community organizations' co-leadership is essential for fostering more effective, sustainable, and ethically sound development of technology in social applications.

\subsection{Rethinking AI4SG: From Tech-Centered to Community-Centered Approaches} 

% the wider phenomenon
The aspiration to employ AI to address complex social issues is not a recent phenomenon. Existing literature connects AI4SG to prior ``tech for good'' waves labeled by different names~\cite{aula_stepping_2023}. Despite limited concrete evidence of social impact and benefits to intended end beneficiaries~\cite{vinuesa_role_2020, pruss_ghosting_2023, ismail_public_2023, Harris2016how, schelenz_information_2022}, AI4SG initiatives continue to capture funders', technology teams', and community organizations' attention due to the ``charisma'' \cite{ames_charisma_2019}, ``magic'' \cite{Lupetti2024unmaking}, and ``enchantment'' \cite{campolo_enchanted_2020} of innovations like AI. In our study, we see how this broader fascination with AI, whether among community organizations, technical teams, or funding agendas, shaped AI4SG partnerships in various ways. 

%\citet{Sims_2017} identified a cyclical pattern of these waves, despite repeated failed attempts to show their effectiveness. \citet{Sims_2017} offer an explanation for this cyclical pattern---failures were framed as a learning process and ended up benefiting those with power---for instance, designers and researchers getting prestigious grants and jobs from these programs. \citet{crawford_atlas_2021} pointed to the , where the fascination with the mysterious capabilities of new technologies captures public attention, attracting funds and power toward techno-solutionism. This echoes lessons learned from previous ICTD programs~\cite{ames_charisma_2019} \deleted{, such as the ``One Laptop per Child,'' as detailed in \textit{The Charisma Machine} by \citet{ames_charisma_2019}}. Prior work in CSCW and HCI like \citet{kapania_because_2022} provide empirical evidence for a sentiment of ``AI authority'' on the ground, where AI is seen to have a legitimized power without requiring evidence about the accuracy and safety of the system. Despite a lack of evidence for its promised effectiveness~\cite{vinuesa_role_2020, pruss_ghosting_2023, ismail_public_2023}, there is a substantial surge in support for AI4SG initiatives, as indicated by the increasing trend highlighted by \citet{shi_artificial_2020}.

% how does it affect stakeholders
%% organizations tendency for innovations
%% outcome 
First, our participants (and their partners and funders) were driven to AI and technology innovations, aspiring to keep pace with advancements in the private sector. However, as shown in our findings, projects often did not deliver what the community organizations had hoped for, while benefiting technology teams, like research labs, with publications and academic prestige. This AI-centric approach succumbs to what is known as the ``law of the instrument'', akin to a hammer searching for a nail. When AI developers and funders exclusively offer AI solutions to organizations, it pre-defines the scope of possibilities available and can lead to missed opportunities where simpler solutions might have been more effective, as has been found in prior work as well~\cite{holzmeyer2021beyond,radhakrishnan_experiments_2021, pruss_ghosting_2023}. Critical perspectives in ICTD have shown that a techno-centric approach without addressing larger societal issues will not result in meaningful, sustained social impact (e.g.~\cite{toyama_geek_2015, ames_charisma_2019, Harris2016how, schelenz_information_2022}).

%% forms of knowledge
Second, even when projects fell short of expectations for deployment, community organization collaborators' recognition of AI developers' expertise remained high and sustained over time. As shown in P15's case, this perception of expertise in technology teams could even reduce community organization members' tendency to take on leadership roles in projects (cf.~\cite{espinoza_big_2021}). We thus contribute an understanding of how discourse around the promise of AI and technical expertise does not just unduly increase expectations of their efficacy, but also has very real material impacts on ownership of AI projects. This increases the urgency with which funders, technology teams, and community organizations must actively combat the notion that AI experts are the only stakeholders well-positioned to steward resources.

%It could also reduce expectations of accountability of AI developers and systems~\cite{ramesh2022platform, kapania_because_2022}, eventually leading to shifting power towards those possessing technological proficiency~\cite{dignazio_data_2020, espinoza_big_2021}, and affecting who gets to have a ``seat at the table" and whose forms of knowledge are privileged in AI development and regulations~\cite{dignazio_data_2020}. %Individuals or organizations possessing valuable insights into social issues are hence marginalized, lacking influence over project selection, solution strategies, and impact evaluation.%, violating the proposed principle of inclusivity in AI development~\cite{un_ai_2023, moon_searching_2023}.
 
%As highlighted in \textit{The Atlas of AI} by \citet{crawford_atlas_2021}, the focus of AI should shift from asking where AI can be applied merely because it can to why it ought to be applied in the first place. 
We argue that AI4SG stakeholders should take a reflexive stance and focus on the social issues at hand, instead of AI, and invest in solutions with demonstrated effectiveness. Instead of \textit{``AI for} social good,'' a more effective approach may be ``social good \textit{with AI}.'' In particular, those who hold the power of shaping funding agendas should consider shifting the focus of funding from only innovative tools to also engaging with social sector organizations or the intended end beneficiaries. For example, this could be accomplished by involving community organizations in the funding design process and allowing them to define the priorities and success metrics of funded projects. Furthermore, many participants in our study pointed out that their projects could not reach deployment because of the short-term nature of their funding. AI4SG funders should consider extending typical funding timelines to acknowledge the complexity and difficulty of aligning technology design with social problems. More broadly, future work in HCI and CSCW could hone in on the power of funders and decision-makers and suggest ways to configure the funding landscape (as suggested in prior work~\cite{saha2022commissioning, Gardner2022ethical}), in order to incentivize addressing the root causes of social problems and engaging with impacted communities meaningfully.

\subsection{Reorienting Power Dynamics in AI4SG Partnerships}

%Extending prior work on how power imbalances affect data systems and data work (e.g.~\cite{miceli2022studying, wang2022whose}) in the non-profit sector (e.g.,~\cite{sandberg2023re, bopp_disempowered_2017}), we discuss how power asymmetries were reflected in various dimensions of AI4SG collaborations.

%% relate to existing framework and the third group represents distinct motivations -- need to read related work to engage with it more. 
First and foremost, funders' agendas, as we examined, had a pervasive influence on AI4SG collaborations. They influenced how partnerships are formed, what problems get prioritized, what solutions are explored, and how long the project lasts (cf.~\cite{Harrington_2019, erete_method_2023}). Prior work also points out that funders' focus on quantitative measurements and data in the evaluation process could also shift mission-driven organizations' focus away from their social missions~\cite{bopp_disempowered_2017}, community-building efforts~\cite{Sum_2023}, and richer qualitative data~\cite{bay2020community, bopp2023}. 

Our participants, especially the Strategists, who did not initiate the projects but were approached by technology teams, were motivated by several external factors to participate. Notably, our participants expressed a need to form partnerships with esteemed institutions, aiming to enhance their organizations' credibility to secure funding. %Another impetus for entering into collaborations is the necessity for access to computing infrastructures, which are costly to acquire and maintain. This speaks to the broader issue of a concentration of computing power within a select few technology companies on a global scale, with non-profit organizations being on the margins of the data economy~\cite{darian2023enacting}. 
Existing data-driven social partnership models highlight social sector organizations' need for resources or learning and responsibility towards a social mission~\cite{susha_achieving_2023, susha_data_2019}. To make \textit{external} power imbalances in AI4SG partnerships explicit~\cite{miceli2022studying}, we suggest extending existing frameworks~\cite{susha_achieving_2023, susha_data_2019} to include the \textit{power dependence} factor, where organizations are also incentivized to collaborate with external partners in response to the interests and priorities of more influential stakeholders such as technology teams and funders.

%% whose goals were prioritized?
Power imbalances were also evident in how projects were evaluated, with goals often reflecting the priorities of technology teams, leading to community organizations' goals being sidelined (cf.~\cite{Kumar2018towards, erete_method_2023}). Technology teams tended to focus on the predictive accuracy of AI systems, which did not always translate to tangible social impact for our participants (cf.~\cite{bingley2023human, ismail_public_2023, bopp_disempowered_2017, bopp2023}). The short-term focus of AI4SG projects further disadvantaged organizations working on complex issues requiring long-term solutions. Additionally, power imbalances within organizations influenced AI adoption, with those closer to the day-to-day implementation holding more realistic views but having less influence on decision-making compared to those more removed from operational realities (cf.~\cite{kawakami_studying_2023, Saha2022towards, Sims_2017}).

% so what: shift power by letting community orgs set goals, don't overpromise, coliberation
It is imperative to reorient relationships with community organizations to challenge power, viewing them not merely as data collectors and annotators but as co-designers and decision-makers~\cite{Sum_2023}. In the subsequent section, we propose the concept of \textit{data co-liberation}~\cite{dignazio_data_2020} as a pathway forward.

\subsection{A Way Forward: ``Data Co-liberation'' instead of ``Data For Good"}
% 3: a way forward: coliberation as a way to re-distribute power
% - relationship-oriented 
% - problem-oriented (might not be AI!)

Community organization members in our study demanded a relationship-first and community-led approach to AI4SG projects. Our participants wanted technology teams to \textit{``come to us first''} and engage with them early, seeking their input before determining the problems to address and the solutions to explore. The underlying rationale and advantage of centering community organizations are clear---community organization collaborators help realize the promise of AI4SG projects effectively and ethically.

% explain what ``data coliberation'' is and why it is a better approach
\textit{Data co-liberation}, as described by \citet{dignazio_data_2020}, embraces pluralism of knowledge and centers the leadership of community organizations and local communities. Beyond viewing them as domain experts and assets~\cite{sambasivan_deskilling_2022, darian2023enacting, miceli2022data} or making their labor visible~\cite{dignazio_data_2020, thakkar2022machine, passi2020making}, \textit{data co-liberation} requires viewing them as leaders and co-designers~\cite{dignazio_data_2020, erete_method_2023, Harrington_2019, Sum_2023}. By sharing agency and power (and money) with community partners, we can shift the focus from the tools to the problems at hand, as well as aligning goals across stakeholders from the beginning~\cite{ismail_public_2023}. Furthermore, as demonstrated in our findings, community organization members provide critical infrastructural and intellectual contributions, ensuring that AI solutions 1) work on the ground, 2) are ethically developed and implemented, especially if they impact vulnerable populations, and 3) are sustainable by emphasizing internal capacity building. Social sector organizations have the \textit{agency} and \textit{capacity} to shape data-driven work to fit broader social missions~\cite{tran2022careful, voida2011homebrew,khovanskaya2020bottom, alvarado2017making,tran2022careful,meng2018grassroots}, and there are many benefits of leveraging community organizations' contextual knowledge and domain expertise into the data project design~\cite{Jung_2022, alvarado2023mobilizing, Dimas_2023}. %Aligned with the assets-based approach to design processes~\cite{kretzmann_assets-based_1996, mathie_clients_2003, wong-villacres_reflections_2021}), seeing community partners as knowledgeable and as equal contributors places the strengths and capacities that individuals and communities already have at the center of research and design. 
Contrary to an implicit assumption of technology teams as the ``saviors'' of ``deficient'' communities, \textit{data co-liberation} recognizes and amplifies the existing strengths within communities.

% privilege hazard and the knowledge deficiency of computer science - it is about their coliberation too. 
Aligned with existing literature on community-based research (e.g., \cite{erete_method_2023, Harrington_2019}), \textit{co-liberation} also means that technology teams actively acknowledge that they come to the table with significant gaps in their knowledge and that there are structural barriers to long-term collaborations~\cite{Liang_2023, Lodato2018institutional}. Our interviews indicate that AI4SG technology teams often lack domain knowledge, contextual understanding of the local communities where they are outsiders~\cite{LeDantec_Fox_2015, Winschiers-Theophilus_2010}, and familiarity with participatory methods and community engagement guidelines~\cite{delgado2023participatory, Unertl_2016, Winschiers-Theophilus_2010}. As the dominant group in data science, they are also subject to ``privilege hazard''~\cite{dignazio_data_2020} that obstructs their understanding of the lived experiences faced by marginalized groups. \textit{Co-liberation} necessitates humility regarding capacities of AI technologies, countering narratives that position technology teams as ``saviors'' and challenging the overarching themes of techno-solutionism~\cite{kapania_because_2022,crawford_atlas_2021, toyama_geek_2015,schelenz_information_2022}. Indeed, prior work has highlighted the challenges that AI developers face when considering ethical guidelines and engaging with user-centered design (e.g.~\cite{wong2023seeing, varanasi2023currently, holstein2019improving}) and that AI developers desire more support for the early stage of ideation and problem formulation~\cite{yildirim2023investigating}. Leveraging and compensating for community organizations' valuable knowledge and co-leadership is one crucial way to support technology teams. 

To facilitate concrete steps towards \textit{co-liberation} in AI4SG initiatives, technology teams ought to first meaningfully engage with existing community-based research and design guidelines (e.g.,~\cite{erete_method_2023, Unertl_2016}) and consider toolkits (e.g.,~\cite{crxlab, Atlanta_Playbook}) that lay out a roadmap for centering community organizations. For example,~\citet{Unertl_2016} clearly lay out ten principles of community-based research in health informatics, which are applicable to the AI4SG contexts. Principle 4---\textit{``Building collaborative partnerships in all research phases. The community is not just included during data collection, but rather is included from problem definition through results dissemination.''}---is particularly aligned with what our participants asked for~\cite[p.70]{Unertl_2016}. The \textit{Atlanta Community Engagement Playbook}~\cite{Atlanta_Playbook} serves as a strong example of a blueprint for meaningful community engagement, providing steps for service providers such as ``listen and learn'' from key members of the community to get to know them, and ``build capacity'' in communities. Recent efforts connecting the participatory design approach and AI4SG, such as Bondi et al.'s PACT framework~\cite{bondi_envisioning_2021} and Lee et al.'s WeBuildAI framework ~\cite{Lee_2019} for collaborative algorithmic design are steps towards \textit{co-liberation}. 

% funding: within the existing funding constraints, share funding
Still, it is clear that the structural constraints that technology teams face in academia and industry present challenges of adopting existing frameworks and toolkits~\cite{deng2023understanding, balayn2023fairness,heger2020all, varanasi2023currently}. However, within these constraints, we also see an opportunity for technology teams to think creatively about what funding opportunities they could pursue and to leverage non-traditional funding structures to enable more meaningful long-term engagement with community organizations~\cite{Sum_2023, erete_method_2023, DIgnazio_2024}. Our interviews reveal that AI4SG funding tends to end before the project can get to the deployment stage due to the complexity of projects, and it privileges technology teams with more institutional credibility. By actively choosing to reject applying for large grants in early stages of the work, the \textit{Data Against Feminicide} project was able to prioritize developing relationships with community partners~\cite{DIgnazio_2024}. As another example,~\citet{erete_method_2023} co-create funding materials and co-apply for grants outside of academic funding sources with their community partner, so that there is no official ``leader'' of their project nor requirements to disseminate their work in academic venues. In both cases, the authors leverage their institutional privilege to distribute funding and power to their community partners as an act of \textit{co-liberation}. They also demonstrate that community organizations' contributions should be formally valued, and that we should avoid leveraging community organizations' expertise and labor without proper compensation and mutual benefits.

Overall, \textit{co-liberation} emphasizes \textit{co}-thinking, \textit{co}-designing, and \textit{co}-existing as people towards a common goal. In arguing for community organizations' co-leadership in AI4SG partnerships, we are advocating for a relationship-first approach in AI development, where community organizations are seen as people who are dedicated to  addressing a social issue, \textit{co-thinkers} to exchange ideas with, and \textit{experts} who should get credits for their ideas. Community organizations' co-leadership is not only essential for fostering more effective, sustainable, and ethical development of technology, it is also important for building meaningful relationships among people.

%% file: 6-Conclusion.tex
\section{Conclusion}

Using \textit{Data Feminism} \cite{dignazio_data_2020} as a grounding framework, we investigated community organization members' perceptions and experiences with AI4SG partnerships. Drawing on 16 semi-structured interviews with staff from community organizations collaborating on AI4SG projects, we reveal power imbalances between community organizations, technology teams, and funders. These power asymmetries influence community organization members' motivations in participating in AI4SG, what the projects prioritize, how they are evaluated, and eventually, the outcomes of projects. Finding that most projects fell short of community organization members' initial expectations, we present their aspirations for success and more impactful partnerships, namely: 1) adopting a relationship-centric approach by involving community organizations in the ideation stage, 2) allowing community organizations to co-lead throughout the project, and 3) investing in technology training for staff. We analyze how a belief in the promise of AI technologies and the cyclical nature of waves of "tech for good" shape AI4SG collaborations. To ensure effective, ethical, and sustainable collaborations on AI4SG projects, we advocate for \textit{data co-liberation} as a grounding principle that shifts power from the hands of the technology teams and other dominant groups to the hands of community organizations and communities that AI4SG initiatives strive to serve. 

In future work, we intend to investigate power relations among other AI4SG stakeholders, especially funding bodies and intended end beneficiaries, whose perspectives are also important but overlooked in the literature. We also see value in case studies of collaborations that have applied the principles of \textit{data co-liberation}, in order to connect the principles to practice.

%% file: main.bbl
%%% -*-BibTeX-*-
%%% Do NOT edit. File created by BibTeX with style
%%% ACM-Reference-Format-Journals [18-Jan-2012].

\begin{thebibliography}{128}

%%% ====================================================================
%%% NOTE TO THE USER: you can override these defaults by providing
%%% customized versions of any of these macros before the \bibliography
%%% command.  Each of them MUST provide its own final punctuation,
%%% except for \shownote{}, \showDOI{}, and \showURL{}.  The latter two
%%% do not use final punctuation, in order to avoid confusing it with
%%% the Web address.
%%%
%%% To suppress output of a particular field, define its macro to expand
%%% to an empty string, or better, \unskip, like this:
%%%
%%% \newcommand{\showDOI}[1]{\unskip}   % LaTeX syntax
%%%
%%% \def \showDOI #1{\unskip}           % plain TeX syntax
%%%
%%% ====================================================================

\ifx \showCODEN    \undefined \def \showCODEN     #1{\unskip}     \fi
\ifx \showDOI      \undefined \def \showDOI       #1{#1}\fi
\ifx \showISBNx    \undefined \def \showISBNx     #1{\unskip}     \fi
\ifx \showISBNxiii \undefined \def \showISBNxiii  #1{\unskip}     \fi
\ifx \showISSN     \undefined \def \showISSN      #1{\unskip}     \fi
\ifx \showLCCN     \undefined \def \showLCCN      #1{\unskip}     \fi
\ifx \shownote     \undefined \def \shownote      #1{#1}          \fi
\ifx \showarticletitle \undefined \def \showarticletitle #1{#1}   \fi
\ifx \showURL      \undefined \def \showURL       {\relax}        \fi
% The following commands are used for tagged output and should be
% invisible to TeX
\providecommand\bibfield[2]{#2}
\providecommand\bibinfo[2]{#2}
\providecommand\natexlab[1]{#1}
\providecommand\showeprint[2][]{arXiv:#2}

\bibitem[Ahmed and Irani(2020)]%
        {Ahmed_2020}
\bibfield{author}{\bibinfo{person}{Alex Ahmed} {and} \bibinfo{person}{Lilly Irani}.} \bibinfo{year}{2020}\natexlab{}.
\newblock \showarticletitle{Feminism as a design methodology}.
\newblock \bibinfo{journal}{\emph{Interactions}} \bibinfo{volume}{27}, \bibinfo{number}{6} (\bibinfo{date}{Nov} \bibinfo{year}{2020}), \bibinfo{pages}{42–45}.
\newblock
\showISSN{1072-5520}
\urldef\tempurl%
\url{https://doi.org/10.1145/3426366}
\showDOI{\tempurl}


\bibitem[Alvarado et~al\mbox{.}({[n.\,d.]})]%
        {Atlanta_Playbook}
\bibfield{author}{\bibinfo{person}{Taranji Alvarado}, \bibinfo{person}{Miriam Asad}, \bibinfo{person}{Patrice Barlow}, \bibinfo{person}{Jr. Chuck~Barlow}, \bibinfo{person}{Al Bartell}, \bibinfo{person}{Terica Black}, \bibinfo{person}{Kelly Brown}, \bibinfo{person}{Sandra Bush}, \bibinfo{person}{Bill Cannon}, \bibinfo{person}{Eric Corbett}, \bibinfo{person}{Kate Diedrick}, \bibinfo{person}{Pamela Flores}, \bibinfo{person}{Nasim Fluker}, \bibinfo{person}{Jhordan Gibbs}, \bibinfo{person}{Lyndon Greene}, \bibinfo{person}{Kendace Hall}, \bibinfo{person}{Gerald Jackson}, \bibinfo{person}{Jr. Graham~Jackson}, \bibinfo{person}{David~“El” Lawrence}, \bibinfo{person}{Christopher~Le Dantec}, \bibinfo{person}{Jr. James~Lewis}, \bibinfo{person}{Becky Nielson}, \bibinfo{person}{Cynthia Parks}, \bibinfo{person}{Lynn Roney}, \bibinfo{person}{Terry Ross}, \bibinfo{person}{Yvonne Wagner}, \bibinfo{person}{Sandra Walker}, {and} \bibinfo{person}{Gwen Weddington}.} \bibinfo{year}{[n.\,d.]}\natexlab{}.
\newblock \bibinfo{booktitle}{\emph{Atlanta Community Engagement Playbook}}.
\newblock
\urldef\tempurl%
\url{http://ourcommunity.is/engaged/}
\showURL{%
\tempurl}
\newblock
\shownote{Created in collaboration with the residents of Ashview Heights, Atlanta University Center, Booker T. Washington, Castleberry Hill, English Avenue, Vine City, and several organizations including the City of Atlanta, Westside Future Fund, Atlanta Housing Authority, and University Choice Neighborhood. Accessed 2024-04-29.}.


\bibitem[Alvarado~Garcia et~al\mbox{.}(2023)]%
        {alvarado2023mobilizing}
\bibfield{author}{\bibinfo{person}{Adriana Alvarado~Garcia}, \bibinfo{person}{Marisol Wong-Villacres}, \bibinfo{person}{Milagros Miceli}, \bibinfo{person}{Benjam{\'\i}n Hern{\'a}ndez}, {and} \bibinfo{person}{Christopher~A Le~Dantec}.} \bibinfo{year}{2023}\natexlab{}.
\newblock \showarticletitle{Mobilizing Social Media Data: Reflections of a Researcher Mediating between Data and Organization}. In \bibinfo{booktitle}{\emph{Proceedings of the 2023 CHI Conference on Human Factors in Computing Systems}}. \bibinfo{pages}{1--19}.
\newblock
\urldef\tempurl%
\url{https://doi.org/10.1145/3544548.3580916}
\showDOI{\tempurl}


\bibitem[Alvarado~Garcia et~al\mbox{.}(2017)]%
        {alvarado2017making}
\bibfield{author}{\bibinfo{person}{Adriana Alvarado~Garcia}, \bibinfo{person}{Alyson~L Young}, {and} \bibinfo{person}{Lynn Dombrowski}.} \bibinfo{year}{2017}\natexlab{}.
\newblock \showarticletitle{On making data actionable: How activists use imperfect data to foster social change for human rights violations in Mexico}.
\newblock \bibinfo{journal}{\emph{Proceedings of the ACM on Human-Computer Interaction}} \bibinfo{volume}{1}, \bibinfo{number}{CSCW} (\bibinfo{year}{2017}), \bibinfo{pages}{1--19}.
\newblock
\urldef\tempurl%
\url{https://doi.org/10.1145/3134654}
\showDOI{\tempurl}


\bibitem[Ames(2019)]%
        {ames_charisma_2019}
\bibfield{author}{\bibinfo{person}{Morgan~G. Ames}.} \bibinfo{year}{2019}\natexlab{}.
\newblock \bibinfo{booktitle}{\emph{The Charisma Machine: The Life, Death, and Legacy of One Laptop per Child}}.
\newblock \bibinfo{publisher}{{MIT} Press}.
\newblock
\showISBNx{978-0-262-53744-5}


\bibitem[Aula and Bowles(2023)]%
        {aula_stepping_2023}
\bibfield{author}{\bibinfo{person}{Ville Aula} {and} \bibinfo{person}{James Bowles}.} \bibinfo{year}{2023}\natexlab{}.
\newblock \showarticletitle{Stepping back from Data and {AI} for Good – current trends and ways forward}.
\newblock \bibinfo{journal}{\emph{Big Data \& Society}} \bibinfo{volume}{10}, \bibinfo{number}{1} (\bibinfo{date}{Dec.} \bibinfo{year}{2023}).
\newblock
\showISSN{2053-9517}
\urldef\tempurl%
\url{https://doi.org/10.1177/20539517231173901}
\showDOI{\tempurl}


\bibitem[Balayn et~al\mbox{.}(2023)]%
        {balayn2023fairness}
\bibfield{author}{\bibinfo{person}{Agathe Balayn}, \bibinfo{person}{Mireia Yurrita}, \bibinfo{person}{Jie Yang}, {and} \bibinfo{person}{Ujwal Gadiraju}.} \bibinfo{year}{2023}\natexlab{}.
\newblock \showarticletitle{``Fairness Toolkits, A Checkbox Culture?'' On the Factors that Fragment Developer Practices in Handling Algorithmic Harms}. In \bibinfo{booktitle}{\emph{Proceedings of the 2023 AAAI/ACM Conference on AI, Ethics, and Society}}. \bibinfo{publisher}{Association for Computing Machinery}, \bibinfo{address}{New York, NY, USA}, \bibinfo{pages}{482--495}.
\newblock
\urldef\tempurl%
\url{https://doi.org/10.1145/3600211.3604674}
\showDOI{\tempurl}


\bibitem[Bardzell(2010)]%
        {Bardzell_2010}
\bibfield{author}{\bibinfo{person}{Shaowen Bardzell}.} \bibinfo{year}{2010}\natexlab{}.
\newblock \showarticletitle{Feminist HCI: taking stock and outlining an agenda for design}. In \bibinfo{booktitle}{\emph{Proceedings of the SIGCHI Conference on Human Factors in Computing Systems}} \emph{(\bibinfo{series}{CHI ’10})}. \bibinfo{publisher}{Association for Computing Machinery}, \bibinfo{address}{New York, NY, USA}, \bibinfo{pages}{1301–1310}.
\newblock
\showISBNx{978-1-60558-929-9}
\urldef\tempurl%
\url{https://doi.org/10.1145/1753326.1753521}
\showDOI{\tempurl}


\bibitem[Bardzell and Bardzell(2011)]%
        {Bardzell_Bardzell_2011}
\bibfield{author}{\bibinfo{person}{Shaowen Bardzell} {and} \bibinfo{person}{Jeffrey Bardzell}.} \bibinfo{year}{2011}\natexlab{}.
\newblock \showarticletitle{Towards a feminist HCI methodology: social science, feminism, and HCI}. In \bibinfo{booktitle}{\emph{Proceedings of the SIGCHI Conference on Human Factors in Computing Systems}} \emph{(\bibinfo{series}{CHI ’11})}. \bibinfo{publisher}{Association for Computing Machinery}, \bibinfo{address}{New York, NY, USA}, \bibinfo{pages}{675–684}.
\newblock
\showISBNx{978-1-4503-0228-9}
\urldef\tempurl%
\url{https://doi.org/10.1145/1978942.1979041}
\showDOI{\tempurl}


\bibitem[Batool et~al\mbox{.}(2021)]%
        {batool2021detecting}
\bibfield{author}{\bibinfo{person}{Amna Batool}, \bibinfo{person}{Kentaro Toyama}, \bibinfo{person}{Tiffany Veinot}, \bibinfo{person}{Beenish Fatima}, {and} \bibinfo{person}{Mustafa Naseem}.} \bibinfo{year}{2021}\natexlab{}.
\newblock \showarticletitle{Detecting Data Falsification by Front-line Development Workers: A Case Study of Vaccination in Pakistan}. In \bibinfo{booktitle}{\emph{Proceedings of the 2021 CHI Conference on Human Factors in Computing Systems}}. \bibinfo{pages}{1--14}.
\newblock
\urldef\tempurl%
\url{https://doi.org/10.1145/3411764.3445630}
\showDOI{\tempurl}


\bibitem[Bay and Yankura~Swacha(2020)]%
        {bay2020community}
\bibfield{author}{\bibinfo{person}{Jennifer~L Bay} {and} \bibinfo{person}{Kathryn Yankura~Swacha}.} \bibinfo{year}{2020}\natexlab{}.
\newblock \showarticletitle{Community-Engaged Research as Enmeshed Practice.}
\newblock \bibinfo{journal}{\emph{Michigan Journal of Community Service Learning}} \bibinfo{volume}{26}, \bibinfo{number}{1} (\bibinfo{year}{2020}), \bibinfo{pages}{121--141}.
\newblock
\urldef\tempurl%
\url{https://doi.org/10.3998/mjcsloa.3239521.0026.108}
\showDOI{\tempurl}


\bibitem[Berendt(2019)]%
        {berendt_ai_2019}
\bibfield{author}{\bibinfo{person}{Bettina Berendt}.} \bibinfo{year}{2019}\natexlab{}.
\newblock \showarticletitle{{AI} for the Common Good?! Pitfalls, challenges, and ethics pen-testing}.
\newblock \bibinfo{journal}{\emph{Paladyn, Journal of Behavioral Robotics}} \bibinfo{volume}{10}, \bibinfo{number}{1} (\bibinfo{date}{Jan.} \bibinfo{year}{2019}), \bibinfo{pages}{44--65}.
\newblock
\showISSN{2081-4836}
\urldef\tempurl%
\url{https://doi.org/10.1515/pjbr-2019-0004}
\showDOI{\tempurl}


\bibitem[Bingley et~al\mbox{.}(2023)]%
        {bingley2023human}
\bibfield{author}{\bibinfo{person}{William~J Bingley}, \bibinfo{person}{Caitlin Curtis}, \bibinfo{person}{Steven Lockey}, \bibinfo{person}{Alina Bialkowski}, \bibinfo{person}{Nicole Gillespie}, \bibinfo{person}{S~Alexander Haslam}, \bibinfo{person}{Ryan~KL Ko}, \bibinfo{person}{Niklas Steffens}, \bibinfo{person}{Janet Wiles}, {and} \bibinfo{person}{Peter Worthy}.} \bibinfo{year}{2023}\natexlab{}.
\newblock \showarticletitle{Where is the human in human-centered AI? Insights from developer priorities and user experiences}.
\newblock \bibinfo{journal}{\emph{Computers in Human Behavior}}  \bibinfo{volume}{141} (\bibinfo{year}{2023}), \bibinfo{pages}{107617}.
\newblock
\urldef\tempurl%
\url{https://doi.org/10.1016/j.chb.2022.107617}
\showDOI{\tempurl}


\bibitem[Bondi et~al\mbox{.}(2021)]%
        {bondi_envisioning_2021}
\bibfield{author}{\bibinfo{person}{Elizabeth Bondi}, \bibinfo{person}{Lily Xu}, \bibinfo{person}{Diana Acosta-Navas}, {and} \bibinfo{person}{Jackson~A. Killian}.} \bibinfo{year}{2021}\natexlab{}.
\newblock \showarticletitle{Envisioning Communities: A Participatory Approach Towards {AI} for Social Good}. In \bibinfo{booktitle}{\emph{Proceedings of the 2021 {AAAI}/{ACM} Conference on {AI}, Ethics, and Society}} (New York, {NY}, {USA}, 2021-07-30) \emph{(\bibinfo{series}{{AIES} '21})}. \bibinfo{publisher}{Association for Computing Machinery}, \bibinfo{pages}{425--436}.
\newblock
\showISBNx{978-1-4503-8473-5}
\urldef\tempurl%
\url{https://doi.org/10.1145/3461702.3462612}
\showDOI{\tempurl}


\bibitem[Boone et~al\mbox{.}(2023)]%
        {boone2023data}
\bibfield{author}{\bibinfo{person}{Ashley Boone}, \bibinfo{person}{Carl Disalvo}, {and} \bibinfo{person}{Christopher~A Le~Dantec}.} \bibinfo{year}{2023}\natexlab{}.
\newblock \showarticletitle{Data Practice for a Politics of Care: Food Assistance as a Site of Careful Data Work}. In \bibinfo{booktitle}{\emph{Proceedings of the 2023 CHI Conference on Human Factors in Computing Systems}}. \bibinfo{pages}{1--13}.
\newblock
\urldef\tempurl%
\url{https://doi.org/10.1145/3544548.3580831}
\showDOI{\tempurl}


\bibitem[Bopp et~al\mbox{.}(2017)]%
        {bopp_disempowered_2017}
\bibfield{author}{\bibinfo{person}{Chris Bopp}, \bibinfo{person}{Ellie Harmon}, {and} \bibinfo{person}{Amy Voida}.} \bibinfo{year}{2017}\natexlab{}.
\newblock \showarticletitle{Disempowered by Data: Nonprofits, Social Enterprises, and the Consequences of Data-Driven Work}. In \bibinfo{booktitle}{\emph{Proceedings of the 2017 {CHI} Conference on Human Factors in Computing Systems}} (Denver Colorado {USA}, 2017-05-02). \bibinfo{publisher}{{ACM}}, \bibinfo{pages}{3608--3619}.
\newblock
\showISBNx{978-1-4503-4655-9}
\urldef\tempurl%
\url{https://doi.org/10.1145/3025453.3025694}
\showDOI{\tempurl}


\bibitem[Bopp and Voida(2023)]%
        {bopp2023}
\bibfield{author}{\bibinfo{person}{Chris Bopp} {and} \bibinfo{person}{Amy Voida}.} \bibinfo{year}{2023}\natexlab{}.
\newblock \showarticletitle{"Showing the Context": A Need for Oligopticonic Information Systems in Homelessness Measurement}.
\newblock \bibinfo{journal}{\emph{Proc. ACM Hum.-Comput. Interact.}} \bibinfo{volume}{7}, \bibinfo{number}{CSCW1}, Article \bibinfo{articleno}{146} (\bibinfo{date}{4} \bibinfo{year}{2023}), \bibinfo{numpages}{24}~pages.
\newblock
\urldef\tempurl%
\url{https://doi.org/10.1145/3579622}
\showDOI{\tempurl}


\bibitem[Bratteteig and Wagner(2016a)]%
        {Bratteteig2016unpacking}
\bibfield{author}{\bibinfo{person}{Tone Bratteteig} {and} \bibinfo{person}{Ina Wagner}.} \bibinfo{year}{2016}\natexlab{a}.
\newblock \showarticletitle{Unpacking the Notion of Participation in Participatory Design}.
\newblock \bibinfo{journal}{\emph{Comput. Supported Coop. Work}} \bibinfo{volume}{25}, \bibinfo{number}{6} (\bibinfo{date}{Dec.} \bibinfo{year}{2016}), \bibinfo{pages}{425–475}.
\newblock
\showISSN{0925-9724}
\urldef\tempurl%
\url{https://doi.org/10.1007/s10606-016-9259-4}
\showDOI{\tempurl}


\bibitem[Bratteteig and Wagner(2016b)]%
        {Bratteteig_Wagner_2016}
\bibfield{author}{\bibinfo{person}{Tone Bratteteig} {and} \bibinfo{person}{Ina Wagner}.} \bibinfo{year}{2016}\natexlab{b}.
\newblock \showarticletitle{What is a participatory design result?}. In \bibinfo{booktitle}{\emph{Proceedings of the 14th Participatory Design Conference: Full papers - Volume 1}} \emph{(\bibinfo{series}{PDC ’16})}. \bibinfo{publisher}{Association for Computing Machinery}, \bibinfo{address}{New York, NY, USA}, \bibinfo{pages}{141–150}.
\newblock
\showISBNx{978-1-4503-4046-5}
\urldef\tempurl%
\url{https://doi.org/10.1145/2940299.2940316}
\showDOI{\tempurl}


\bibitem[Braun and Clarke(2006)]%
        {braun_using_2006}
\bibfield{author}{\bibinfo{person}{Virginia Braun} {and} \bibinfo{person}{Victoria Clarke}.} \bibinfo{year}{2006}\natexlab{}.
\newblock \showarticletitle{Using thematic analysis in psychology}.
\newblock \bibinfo{journal}{\emph{Qualitative Research in Psychology}} \bibinfo{volume}{3}, \bibinfo{number}{2} (\bibinfo{year}{2006}), \bibinfo{pages}{77--101}.
\newblock
\showISSN{1478-0887, 1478-0895}
\urldef\tempurl%
\url{https://doi.org/10.1191/1478088706qp063oa}
\showDOI{\tempurl}


\bibitem[Braun and Clarke(2019)]%
        {Braun_Clarke_2019}
\bibfield{author}{\bibinfo{person}{Virginia Braun} {and} \bibinfo{person}{Victoria Clarke}.} \bibinfo{year}{2019}\natexlab{}.
\newblock \showarticletitle{Reflecting on reflexive thematic analysis}.
\newblock \bibinfo{journal}{\emph{Qualitative Research in Sport, Exercise and Health}} \bibinfo{volume}{11}, \bibinfo{number}{4} (\bibinfo{date}{Aug.} \bibinfo{year}{2019}), \bibinfo{pages}{589–597}.
\newblock
\showISSN{2159-676X, 2159-6778}
\urldef\tempurl%
\url{https://doi.org/10.1080/2159676X.2019.1628806}
\showDOI{\tempurl}


\bibitem[Brown and Mickelson(2019)]%
        {brown2019some}
\bibfield{author}{\bibinfo{person}{Suzana Brown} {and} \bibinfo{person}{Alan Mickelson}.} \bibinfo{year}{2019}\natexlab{}.
\newblock \showarticletitle{Why some well-planned and community-based ICTD interventions fail}.
\newblock \bibinfo{journal}{\emph{Information Technologies \& International Development}}  \bibinfo{volume}{15} (\bibinfo{year}{2019}), \bibinfo{pages}{13}.
\newblock


\bibitem[Campolo and Crawford(2020)]%
        {campolo_enchanted_2020}
\bibfield{author}{\bibinfo{person}{Alexander Campolo} {and} \bibinfo{person}{Kate Crawford}.} \bibinfo{year}{2020}\natexlab{}.
\newblock \showarticletitle{Enchanted Determinism: Power without Responsibility in Artificial Intelligence}.
\newblock \bibinfo{journal}{\emph{Engaging Science, Technology, and Society}}  \bibinfo{volume}{6} (\bibinfo{year}{2020}), \bibinfo{pages}{1--19}.
\newblock
\showISSN{2413-8053}
\urldef\tempurl%
\url{https://doi.org/10.17351/ests2020.277}
\showDOI{\tempurl}


\bibitem[Charmaz(2012)]%
        {charmaz_constructing_2012}
\bibfield{author}{\bibinfo{person}{Kathy Charmaz}.} \bibinfo{year}{2012}\natexlab{}.
\newblock \bibinfo{booktitle}{\emph{Constructing grounded theory: a practical guide through qualitative analysis} (\bibinfo{edition}{repr} ed.)}.
\newblock \bibinfo{publisher}{Sage}.
\newblock
\showISBNx{978-0-7619-7352-2 978-0-7619-7353-9}


\bibitem[Chen et~al\mbox{.}(2023)]%
        {chen2023maintainers}
\bibfield{author}{\bibinfo{person}{Yuchen Chen}, \bibinfo{person}{Yuling Sun}, {and} \bibinfo{person}{Silvia Lindtner}.} \bibinfo{year}{2023}\natexlab{}.
\newblock \showarticletitle{Maintainers of Stability: The Labor of China’s Data-Driven Governance and Dynamic Zero-COVID}. In \bibinfo{booktitle}{\emph{Proceedings of the 2023 CHI Conference on Human Factors in Computing Systems}}. \bibinfo{pages}{1--16}.
\newblock
\urldef\tempurl%
\url{https://doi.org/10.1145/3544548.3581299}
\showDOI{\tempurl}


\bibitem[Cowls(2021)]%
        {cowls_ai_2021}
\bibfield{author}{\bibinfo{person}{Josh Cowls}.} \bibinfo{year}{2021}\natexlab{}.
\newblock \showarticletitle{‘{AI} for Social Good’: Whose Good and Who’s Good? Introduction to the Special Issue on Artificial Intelligence for Social Good}.
\newblock \bibinfo{journal}{\emph{Philosophy \& Technology}} \bibinfo{volume}{34}, \bibinfo{number}{1} (\bibinfo{date}{Nov.} \bibinfo{year}{2021}), \bibinfo{pages}{1--5}.
\newblock
\showISSN{2210-5441}
\urldef\tempurl%
\url{https://doi.org/10.1007/s13347-021-00466-3}
\showDOI{\tempurl}


\bibitem[Crawford(2021)]%
        {crawford_atlas_2021}
\bibfield{author}{\bibinfo{person}{Kate Crawford}.} \bibinfo{year}{2021}\natexlab{}.
\newblock \bibinfo{booktitle}{\emph{The Atlas of {AI}: Power, Politics, and the Planetary Costs of Artificial Intelligence}}.
\newblock \bibinfo{publisher}{Yale University Press}.
\newblock
\showISBNx{978-0-300-20957-0}


\bibitem[Darian et~al\mbox{.}(2023)]%
        {darian2023enacting}
\bibfield{author}{\bibinfo{person}{Shiva Darian}, \bibinfo{person}{Aarjav Chauhan}, \bibinfo{person}{Ricky Marton}, \bibinfo{person}{Janet Ruppert}, \bibinfo{person}{Kathleen Anderson}, \bibinfo{person}{Ryan Clune}, \bibinfo{person}{Madeline Cupchak}, \bibinfo{person}{Max Gannett}, \bibinfo{person}{Joel Holton}, \bibinfo{person}{Elizabeth Kamas}, \bibinfo{person}{Jason Kibozi-Yocka}, \bibinfo{person}{Devin Mauro-Gallegos}, \bibinfo{person}{Simon Naylor}, \bibinfo{person}{Meghan O'Malley}, \bibinfo{person}{Mehul Patel}, \bibinfo{person}{Jack Sandberg}, \bibinfo{person}{Troy Siegler}, \bibinfo{person}{Ryan Tate}, \bibinfo{person}{Abigil Temtim}, \bibinfo{person}{Samantha Whaley}, {and} \bibinfo{person}{Amy Voida}.} \bibinfo{year}{2023}\natexlab{}.
\newblock \showarticletitle{Enacting Data Feminism in Advocacy Data Work}.
\newblock \bibinfo{journal}{\emph{Proceedings of the ACM on Human-Computer Interaction}} \bibinfo{volume}{7}, \bibinfo{number}{CSCW1} (\bibinfo{year}{2023}), \bibinfo{pages}{1--28}.
\newblock
\urldef\tempurl%
\url{https://doi.org/10.1145/3579480}
\showDOI{\tempurl}


\bibitem[Dearden and Kleine(2018)]%
        {dearden2018minimum}
\bibfield{author}{\bibinfo{person}{Andy Dearden} {and} \bibinfo{person}{Dorothea Kleine}.} \bibinfo{year}{2018}\natexlab{}.
\newblock \showarticletitle{Minimum ethical standards for ICTD/ICT4D research}.
\newblock \bibinfo{journal}{\emph{C3RI, Sheffield Hallam University}} (\bibinfo{year}{2018}).
\newblock


\bibitem[Delgado et~al\mbox{.}(2023)]%
        {delgado2023participatory}
\bibfield{author}{\bibinfo{person}{Fernando Delgado}, \bibinfo{person}{Stephen Yang}, \bibinfo{person}{Michael Madaio}, {and} \bibinfo{person}{Qian Yang}.} \bibinfo{year}{2023}\natexlab{}.
\newblock \showarticletitle{The participatory turn in ai design: Theoretical foundations and the current state of practice}. In \bibinfo{booktitle}{\emph{Proceedings of the 3rd ACM Conference on Equity and Access in Algorithms, Mechanisms, and Optimization}}. \bibinfo{pages}{1--23}.
\newblock
\urldef\tempurl%
\url{https://doi.org/10.1145/3617694.3623261}
\showDOI{\tempurl}


\bibitem[Dell et~al\mbox{.}(2015)]%
        {Dell2015paper}
\bibfield{author}{\bibinfo{person}{Nicola Dell}, \bibinfo{person}{Trevor Perrier}, \bibinfo{person}{Neha Kumar}, \bibinfo{person}{Mitchell Lee}, \bibinfo{person}{Rachel Powers}, {and} \bibinfo{person}{Gaetano Borriello}.} \bibinfo{year}{2015}\natexlab{}.
\newblock \showarticletitle{Paper-Digital Workflows in Global Development Organizations}. In \bibinfo{booktitle}{\emph{Proceedings of the 18th ACM Conference on Computer Supported Cooperative Work \& Social Computing}} \emph{(\bibinfo{series}{CSCW ’15})}. \bibinfo{publisher}{Association for Computing Machinery}, \bibinfo{address}{New York, NY, USA}, \bibinfo{pages}{1659–1669}.
\newblock
\showISBNx{978-1-4503-2922-4}
\urldef\tempurl%
\url{https://doi.org/10.1145/2675133.2675145}
\showDOI{\tempurl}


\bibitem[Deng et~al\mbox{.}(2023)]%
        {deng2023understanding}
\bibfield{author}{\bibinfo{person}{Wesley~Hanwen Deng}, \bibinfo{person}{Boyuan Guo}, \bibinfo{person}{Alicia Devrio}, \bibinfo{person}{Hong Shen}, \bibinfo{person}{Motahhare Eslami}, {and} \bibinfo{person}{Kenneth Holstein}.} \bibinfo{year}{2023}\natexlab{}.
\newblock \showarticletitle{Understanding Practices, Challenges, and Opportunities for User-Engaged Algorithm Auditing in Industry Practice}. In \bibinfo{booktitle}{\emph{Proceedings of the 2023 CHI Conference on Human Factors in Computing Systems}}. \bibinfo{pages}{1--18}.
\newblock
\urldef\tempurl%
\url{https://doi.org/10.1145/3544548.3581026}
\showDOI{\tempurl}


\bibitem[Deng et~al\mbox{.}(2022)]%
        {deng2022exploring}
\bibfield{author}{\bibinfo{person}{Wesley~Hanwen Deng}, \bibinfo{person}{Manish Nagireddy}, \bibinfo{person}{Michelle Seng~Ah Lee}, \bibinfo{person}{Jatinder Singh}, \bibinfo{person}{Zhiwei~Steven Wu}, \bibinfo{person}{Kenneth Holstein}, {and} \bibinfo{person}{Haiyi Zhu}.} \bibinfo{year}{2022}\natexlab{}.
\newblock \showarticletitle{Exploring how machine learning practitioners (try to) use fairness toolkits}. In \bibinfo{booktitle}{\emph{Proceedings of the 2022 ACM Conference on Fairness, Accountability, and Transparency}}. \bibinfo{pages}{473--484}.
\newblock
\urldef\tempurl%
\url{https://doi.org/10.1145/3531146.3533113}
\showDOI{\tempurl}


\bibitem[D'Ignazio(2024)]%
        {DIgnazio_2024}
\bibfield{author}{\bibinfo{person}{Catherine D'Ignazio}.} \bibinfo{year}{2024}\natexlab{}.
\newblock \showarticletitle{Co-designing for Restorative/Transformative Data Science}.
\newblock In \bibinfo{booktitle}{\emph{Counting Feminicide: Data Feminism in Action}}. \bibinfo{publisher}{The MIT Press}, Chapter~7, \bibinfo{pages}{0}.
\newblock
\showISBNx{978-0-262-37801-7}
\urldef\tempurl%
\url{https://doi.org/10.7551/mitpress/14671.003.0013}
\showDOI{\tempurl}


\bibitem[D'Ignazio and Klein(2020)]%
        {dignazio_data_2020}
\bibfield{author}{\bibinfo{person}{Catherine D'Ignazio} {and} \bibinfo{person}{Lauren~F. Klein}.} \bibinfo{year}{2020}\natexlab{}.
\newblock \bibinfo{booktitle}{\emph{Data Feminism}}.
\newblock \bibinfo{publisher}{{MIT} Press}.
\newblock
\showISBNx{978-0-262-04400-4}


\bibitem[Dimas et~al\mbox{.}(2023)]%
        {Dimas_2023}
\bibfield{author}{\bibinfo{person}{Geri~Louise Dimas}, \bibinfo{person}{Lauri Goldkind}, {and} \bibinfo{person}{Renata Konrad}.} \bibinfo{year}{2023}\natexlab{}.
\newblock \showarticletitle{Big ideas, small data: Opportunities and challenges for data science and the social services sector}.
\newblock \bibinfo{journal}{\emph{Big Data \& Society}} \bibinfo{volume}{10}, \bibinfo{number}{1} (\bibinfo{date}{Jan.} \bibinfo{year}{2023}), \bibinfo{pages}{20539517231171051}.
\newblock
\showISSN{2053-9517}
\urldef\tempurl%
\url{https://doi.org/10.1177/20539517231171051}
\showDOI{\tempurl}


\bibitem[D’Ignazio and F.~Klein(2020)]%
        {DIgnazio_Klein_2020}
\bibfield{author}{\bibinfo{person}{Catherine D’Ignazio} {and} \bibinfo{person}{Lauren F.~Klein}.} \bibinfo{year}{2020}\natexlab{}.
\newblock \showarticletitle{Seven intersectional feminist principles for equitable and actionable COVID-19 data}.
\newblock \bibinfo{journal}{\emph{Big Data \& Society}} \bibinfo{volume}{7}, \bibinfo{number}{2} (\bibinfo{date}{July} \bibinfo{year}{2020}).
\newblock
\showISSN{2053-9517}
\urldef\tempurl%
\url{https://doi.org/10.1177/2053951720942544}
\showDOI{\tempurl}


\bibitem[Elsden et~al\mbox{.}(2019)]%
        {elsden2019sorting}
\bibfield{author}{\bibinfo{person}{Chris Elsden}, \bibinfo{person}{Kate Symons}, \bibinfo{person}{Raluca Bunduchi}, \bibinfo{person}{Chris Speed}, {and} \bibinfo{person}{John Vines}.} \bibinfo{year}{2019}\natexlab{}.
\newblock \showarticletitle{Sorting out valuation in the charity shop: Designing for data-driven innovation through value translation}.
\newblock \bibinfo{journal}{\emph{Proceedings of the ACM on Human-Computer Interaction}} \bibinfo{volume}{3}, \bibinfo{number}{CSCW} (\bibinfo{year}{2019}), \bibinfo{pages}{1--25}.
\newblock
\urldef\tempurl%
\url{https://doi.org/10.1145/3359211}
\showDOI{\tempurl}


\bibitem[Erete et~al\mbox{.}(2023)]%
        {erete_method_2023}
\bibfield{author}{\bibinfo{person}{Sheena Erete}, \bibinfo{person}{Yolanda Rankin}, {and} \bibinfo{person}{Jakita Thomas}.} \bibinfo{year}{2023}\natexlab{}.
\newblock \showarticletitle{A Method to the Madness: Applying an Intersectional Analysis of Structural Oppression and Power in {HCI} and Design}.
\newblock \bibinfo{journal}{\emph{{ACM} Transactions on Computer-Human Interaction}} \bibinfo{volume}{30}, \bibinfo{number}{2} (\bibinfo{year}{2023}), \bibinfo{pages}{24:1--24:45}.
\newblock
\showISSN{1073-0516}
\urldef\tempurl%
\url{https://doi.org/10.1145/3507695}
\showDOI{\tempurl}


\bibitem[Erete et~al\mbox{.}(2016)]%
        {erete_storytelling_2016}
\bibfield{author}{\bibinfo{person}{Sheena Erete}, \bibinfo{person}{Emily Ryou}, \bibinfo{person}{Geoff Smith}, \bibinfo{person}{Khristina~Marie Fassett}, {and} \bibinfo{person}{Sarah Duda}.} \bibinfo{year}{2016}\natexlab{}.
\newblock \showarticletitle{Storytelling with Data: Examining the Use of Data by Non-Profit Organizations}. In \bibinfo{booktitle}{\emph{Proceedings of the 19th {ACM} Conference on Computer-Supported Cooperative Work \& Social Computing}} (New York, {NY}, {USA}, 2016-02-27) \emph{(\bibinfo{series}{{CSCW} '16})}. \bibinfo{publisher}{Association for Computing Machinery}, \bibinfo{pages}{1273--1283}.
\newblock
\showISBNx{978-1-4503-3592-8}
\urldef\tempurl%
\url{https://doi.org/10.1145/2818048.2820068}
\showDOI{\tempurl}


\bibitem[Espinoza and Aronczyk(2021)]%
        {espinoza_big_2021}
\bibfield{author}{\bibinfo{person}{Maria~I Espinoza} {and} \bibinfo{person}{Melissa Aronczyk}.} \bibinfo{year}{2021}\natexlab{}.
\newblock \showarticletitle{Big data for climate action or climate action for big data?}
\newblock \bibinfo{journal}{\emph{Big Data \& Society}} \bibinfo{volume}{8}, \bibinfo{number}{1} (\bibinfo{date}{Jan.} \bibinfo{year}{2021}), \bibinfo{pages}{2053951720982032}.
\newblock
\showISSN{2053-9517}
\urldef\tempurl%
\url{https://doi.org/10.1177/2053951720982032}
\showDOI{\tempurl}
\newblock
\shownote{Publisher: {SAGE} Publications Ltd}.


\bibitem[Eubanks(2018)]%
        {eubanks_automating_2018}
\bibfield{author}{\bibinfo{person}{Virginia Eubanks}.} \bibinfo{year}{2018}\natexlab{}.
\newblock \bibinfo{booktitle}{\emph{Automating Inequality: How High-Tech Tools Profile, Police, and Punish the Poor}}.
\newblock \bibinfo{publisher}{St. Martin's Publishing Group}.
\newblock
\showISBNx{978-1-4668-8596-7}


\bibitem[Fox et~al\mbox{.}(2023)]%
        {fox2023patchwork}
\bibfield{author}{\bibinfo{person}{Sarah~E Fox}, \bibinfo{person}{Samantha Shorey}, \bibinfo{person}{Esther~Y Kang}, \bibinfo{person}{Dominique Montiel~Valle}, {and} \bibinfo{person}{Estefania Rodriguez}.} \bibinfo{year}{2023}\natexlab{}.
\newblock \showarticletitle{Patchwork: the hidden, human labor of AI integration within essential work}.
\newblock \bibinfo{journal}{\emph{Proceedings of the ACM on Human-Computer Interaction}} \bibinfo{volume}{7}, \bibinfo{number}{CSCW1} (\bibinfo{year}{2023}), \bibinfo{pages}{1--20}.
\newblock
\urldef\tempurl%
\url{https://doi.org/10.1145/3579514}
\showDOI{\tempurl}


\bibitem[Gardner et~al\mbox{.}(2022)]%
        {Gardner2022ethical}
\bibfield{author}{\bibinfo{person}{Allison Gardner}, \bibinfo{person}{Adam~Leon Smith}, \bibinfo{person}{Adam Steventon}, \bibinfo{person}{Ellen Coughlan}, {and} \bibinfo{person}{Marie Oldfield}.} \bibinfo{year}{2022}\natexlab{}.
\newblock \showarticletitle{Ethical funding for trustworthy AI: proposals to address the responsibilities of funders to ensure that projects adhere to trustworthy AI practice}.
\newblock \bibinfo{journal}{\emph{AI and Ethics}} \bibinfo{volume}{2}, \bibinfo{number}{2} (\bibinfo{date}{May} \bibinfo{year}{2022}), \bibinfo{pages}{277–291}.
\newblock
\showISSN{2730-5961}
\urldef\tempurl%
\url{https://doi.org/10.1007/s43681-021-00069-w}
\showDOI{\tempurl}


\bibitem[Gray and Suri(2019)]%
        {gray_ghost_2019}
\bibfield{author}{\bibinfo{person}{Mary~L. Gray} {and} \bibinfo{person}{Siddharth Suri}.} \bibinfo{year}{2019}\natexlab{}.
\newblock \bibinfo{booktitle}{\emph{Ghost Work: How to Stop Silicon Valley from Building a New Global Underclass}}.
\newblock \bibinfo{publisher}{Houghton Mifflin Harcourt}.
\newblock
\showISBNx{978-1-328-56624-9}


\bibitem[Harrington et~al\mbox{.}(2019)]%
        {Harrington_2019}
\bibfield{author}{\bibinfo{person}{Christina Harrington}, \bibinfo{person}{Sheena Erete}, {and} \bibinfo{person}{Anne~Marie Piper}.} \bibinfo{year}{2019}\natexlab{}.
\newblock \showarticletitle{Deconstructing Community-Based Collaborative Design: Towards More Equitable Participatory Design Engagements}.
\newblock \bibinfo{journal}{\emph{Proceedings of the ACM on Human-Computer Interaction}} \bibinfo{volume}{3}, \bibinfo{number}{CSCW} (\bibinfo{date}{Nov.} \bibinfo{year}{2019}), \bibinfo{pages}{216:1--216:25}.
\newblock
\urldef\tempurl%
\url{https://doi.org/10.1145/3359318}
\showDOI{\tempurl}


\bibitem[Harris(2016)]%
        {Harris2016how}
\bibfield{author}{\bibinfo{person}{Roger~W. Harris}.} \bibinfo{year}{2016}\natexlab{}.
\newblock \showarticletitle{How ICT4D Research Fails the Poor}.
\newblock \bibinfo{journal}{\emph{Information Technology for Development}} \bibinfo{volume}{22}, \bibinfo{number}{1} (\bibinfo{date}{Jan.} \bibinfo{year}{2016}), \bibinfo{pages}{177–192}.
\newblock
\showISSN{0268-1102}
\urldef\tempurl%
\url{https://doi.org/10.1080/02681102.2015.1018115}
\showDOI{\tempurl}


\bibitem[Hartikainen et~al\mbox{.}(2022)]%
        {hartikainen2022human}
\bibfield{author}{\bibinfo{person}{Maria Hartikainen}, \bibinfo{person}{Kaisa V{\"a}{\"a}n{\"a}nen}, \bibinfo{person}{Anu Lehti{\"o}}, \bibinfo{person}{Saara Ala-Luopa}, {and} \bibinfo{person}{Thomas Olsson}.} \bibinfo{year}{2022}\natexlab{}.
\newblock \showarticletitle{Human-Centered AI Design in Reality: A Study of Developer Companies’ Practices: A study of Developer Companies’ Practices}. In \bibinfo{booktitle}{\emph{Nordic human-computer interaction conference}}. \bibinfo{pages}{1--11}.
\newblock
\urldef\tempurl%
\url{https://doi.org/10.1145/3546155.3546677}
\showDOI{\tempurl}


\bibitem[Hayes(2011)]%
        {Hayes_2011}
\bibfield{author}{\bibinfo{person}{Gillian~R. Hayes}.} \bibinfo{year}{2011}\natexlab{}.
\newblock \showarticletitle{The relationship of action research to human-computer interaction}.
\newblock \bibinfo{journal}{\emph{ACM Transactions on Computer-Human Interaction}} \bibinfo{volume}{18}, \bibinfo{number}{3} (\bibinfo{date}{Aug.} \bibinfo{year}{2011}), \bibinfo{pages}{15:1--15:20}.
\newblock
\showISSN{1073-0516}
\urldef\tempurl%
\url{https://doi.org/10.1145/1993060.1993065}
\showDOI{\tempurl}


\bibitem[Heger et~al\mbox{.}(2020)]%
        {heger2020all}
\bibfield{author}{\bibinfo{person}{Amy Heger}, \bibinfo{person}{Samir Passi}, {and} \bibinfo{person}{Mihaela Vorvoreanu}.} \bibinfo{year}{2020}\natexlab{}.
\newblock \showarticletitle{All the tools, none of the motivation: organizational culture and barriers to responsible AI work}.
\newblock \bibinfo{journal}{\emph{Cultures in AI}} (\bibinfo{year}{2020}).
\newblock


\bibitem[Heger et~al\mbox{.}(2022)]%
        {heger2022understanding}
\bibfield{author}{\bibinfo{person}{Amy~K Heger}, \bibinfo{person}{Liz~B Marquis}, \bibinfo{person}{Mihaela Vorvoreanu}, \bibinfo{person}{Hanna Wallach}, {and} \bibinfo{person}{Jennifer Wortman~Vaughan}.} \bibinfo{year}{2022}\natexlab{}.
\newblock \showarticletitle{Understanding Machine Learning Practitioners' Data Documentation Perceptions, Needs, Challenges, and Desiderata}.
\newblock \bibinfo{journal}{\emph{Proceedings of the ACM on Human-Computer Interaction}} \bibinfo{volume}{6}, \bibinfo{number}{CSCW2} (\bibinfo{year}{2022}), \bibinfo{pages}{1--29}.
\newblock
\urldef\tempurl%
\url{https://doi.org/10.1145/3555760}
\showDOI{\tempurl}


\bibitem[Holstein et~al\mbox{.}(2019)]%
        {holstein2019improving}
\bibfield{author}{\bibinfo{person}{Kenneth Holstein}, \bibinfo{person}{Jennifer Wortman~Vaughan}, \bibinfo{person}{Hal Daum{\'e}~III}, \bibinfo{person}{Miro Dudik}, {and} \bibinfo{person}{Hanna Wallach}.} \bibinfo{year}{2019}\natexlab{}.
\newblock \showarticletitle{Improving fairness in machine learning systems: What do industry practitioners need?}. In \bibinfo{booktitle}{\emph{Proceedings of the 2019 CHI conference on human factors in computing systems}}. \bibinfo{pages}{1--16}.
\newblock
\urldef\tempurl%
\url{https://doi.org/10.1145/3290605.3300830}
\showDOI{\tempurl}


\bibitem[Holzmeyer(2021)]%
        {holzmeyer2021beyond}
\bibfield{author}{\bibinfo{person}{Cheryl Holzmeyer}.} \bibinfo{year}{2021}\natexlab{}.
\newblock \showarticletitle{Beyond ‘AI for Social Good’(AI4SG): social transformations—not tech-fixes—for health equity}.
\newblock \bibinfo{journal}{\emph{Interdisciplinary Science Reviews}} \bibinfo{volume}{46}, \bibinfo{number}{1-2} (\bibinfo{year}{2021}), \bibinfo{pages}{94--125}.
\newblock
\urldef\tempurl%
\url{https://doi.org/10.1080/03080188.2020.1840221}
\showDOI{\tempurl}


\bibitem[Hou and Wang(2017)]%
        {hou2017hacking}
\bibfield{author}{\bibinfo{person}{Youyang Hou} {and} \bibinfo{person}{Dakuo Wang}.} \bibinfo{year}{2017}\natexlab{}.
\newblock \showarticletitle{Hacking with NPOs: collaborative analytics and broker roles in civic data hackathons}.
\newblock \bibinfo{journal}{\emph{Proceedings of the ACM on Human-Computer Interaction}} \bibinfo{volume}{1}, \bibinfo{number}{CSCW} (\bibinfo{year}{2017}), \bibinfo{pages}{1--16}.
\newblock


\bibitem[Ismail and Kumar(2021)]%
        {ismail_ai_2021}
\bibfield{author}{\bibinfo{person}{Azra Ismail} {and} \bibinfo{person}{Neha Kumar}.} \bibinfo{year}{2021}\natexlab{}.
\newblock \showarticletitle{{AI} in Global Health: The View from the Front Lines}. In \bibinfo{booktitle}{\emph{Proceedings of the 2021 {CHI} Conference on Human Factors in Computing Systems}} (New York, {NY}, {USA}, 2021-05-07) \emph{(\bibinfo{series}{{CHI} '21})}. \bibinfo{publisher}{Association for Computing Machinery}, \bibinfo{pages}{1--21}.
\newblock
\showISBNx{978-1-4503-8096-6}
\urldef\tempurl%
\url{https://doi.org/10.1145/3411764.3445130}
\showDOI{\tempurl}


\bibitem[Ismail et~al\mbox{.}(2023)]%
        {ismail_public_2023}
\bibfield{author}{\bibinfo{person}{Azra Ismail}, \bibinfo{person}{Divy Thakkar}, \bibinfo{person}{Neha Madhiwalla}, {and} \bibinfo{person}{Neha Kumar}.} \bibinfo{year}{2023}\natexlab{}.
\newblock \showarticletitle{Public Health Calls for/with {AI}: An Ethnographic Perspective}.
\newblock \bibinfo{journal}{\emph{Proceedings of the {ACM} on Human-Computer Interaction}}  \bibinfo{volume}{7} (\bibinfo{date}{9} \bibinfo{year}{2023}), \bibinfo{pages}{1--26}.
\newblock
Issue {CSCW}2.
\showISSN{2573-0142}
\urldef\tempurl%
\url{https://doi.org/10.1145/3610203}
\showDOI{\tempurl}


\bibitem[Jung et~al\mbox{.}(2022)]%
        {Jung_2022}
\bibfield{author}{\bibinfo{person}{Ju~Yeon Jung}, \bibinfo{person}{Tom Steinberger}, \bibinfo{person}{John~L. King}, {and} \bibinfo{person}{Mark~S. Ackerman}.} \bibinfo{year}{2022}\natexlab{}.
\newblock \showarticletitle{How Domain Experts Work with Data: Situating Data Science in the Practices and Settings of Craftwork}.
\newblock \bibinfo{journal}{\emph{Proceedings of the ACM on Human-Computer Interaction}} \bibinfo{volume}{6}, \bibinfo{number}{CSCW1} (\bibinfo{date}{April} \bibinfo{year}{2022}), \bibinfo{pages}{58:1--58:29}.
\newblock
\urldef\tempurl%
\url{https://doi.org/10.1145/3512905}
\showDOI{\tempurl}


\bibitem[Kapania et~al\mbox{.}(2022)]%
        {kapania_because_2022}
\bibfield{author}{\bibinfo{person}{Shivani Kapania}, \bibinfo{person}{Oliver Siy}, \bibinfo{person}{Gabe Clapper}, \bibinfo{person}{Azhagu~Meena Sp}, {and} \bibinfo{person}{Nithya Sambasivan}.} \bibinfo{year}{2022}\natexlab{}.
\newblock \showarticletitle{”Because {AI} is 100\% right and safe”: User Attitudes and Sources of {AI} Authority in India}. In \bibinfo{booktitle}{\emph{{CHI} Conference on Human Factors in Computing Systems}} (New Orleans {LA} {USA}, 2022-04-27). \bibinfo{publisher}{{ACM}}, \bibinfo{pages}{1--18}.
\newblock
\showISBNx{978-1-4503-9157-3}
\urldef\tempurl%
\url{https://doi.org/10.1145/3491102.3517533}
\showDOI{\tempurl}


\bibitem[Kawakami et~al\mbox{.}(2024)]%
        {kawakami_studying_2023}
\bibfield{author}{\bibinfo{person}{Anna Kawakami}, \bibinfo{person}{Amanda Coston}, \bibinfo{person}{Hoda Heidari}, \bibinfo{person}{Kenneth Holstein}, {and} \bibinfo{person}{Haiyi Zhu}.} \bibinfo{year}{2024}\natexlab{}.
\newblock \bibinfo{title}{Studying Up Public Sector AI: How Networks of Power Relations Shape Agency Decisions Around AI Design and Use}.
\newblock
\newblock
\showeprint[arxiv]{2405.12458}~[cs.HC]
\urldef\tempurl%
\url{https://arxiv.org/abs/2405.12458}
\showURL{%
\tempurl}


\bibitem[Khovanskaya et~al\mbox{.}(2020)]%
        {khovanskaya2020bottom}
\bibfield{author}{\bibinfo{person}{Vera Khovanskaya}, \bibinfo{person}{Phoebe Sengers}, {and} \bibinfo{person}{Lynn Dombrowski}.} \bibinfo{year}{2020}\natexlab{}.
\newblock \showarticletitle{Bottom-Up organizing with tools from on high: Understanding the data practices of labor organizers}. In \bibinfo{booktitle}{\emph{Proceedings of the 2020 CHI Conference on Human Factors in Computing Systems}}. \bibinfo{pages}{1--13}.
\newblock
\urldef\tempurl%
\url{https://doi.org/10.1145/3313831.3376185}
\showDOI{\tempurl}


\bibitem[Kumar and Dell(2018)]%
        {Kumar2018towards}
\bibfield{author}{\bibinfo{person}{Neha Kumar} {and} \bibinfo{person}{Nicola Dell}.} \bibinfo{year}{2018}\natexlab{}.
\newblock \showarticletitle{Towards Informed Practice in HCI for Development}.
\newblock \bibinfo{journal}{\emph{Proc. ACM Hum.-Comput. Interact.}} \bibinfo{volume}{2}, \bibinfo{number}{CSCW} (\bibinfo{date}{Nov.} \bibinfo{year}{2018}), \bibinfo{pages}{99:1--99:20}.
\newblock
\urldef\tempurl%
\url{https://doi.org/10.1145/3274368}
\showDOI{\tempurl}


\bibitem[Lab(2024)]%
        {crxlab}
\bibfield{author}{\bibinfo{person}{Creative~Reaction Lab}.} \bibinfo{year}{2024}\natexlab{}.
\newblock \bibinfo{title}{Equity-Centered Community Design Field Guide}.
\newblock
\newblock
\urldef\tempurl%
\url{https://crxlab.org/shop/p/field-guide-equity-centered-community-design}
\showURL{%
\tempurl}
\newblock
\shownote{Accessed 2024-01-04}.


\bibitem[Le~Dantec(2016)]%
        {LeDantec_2016}
\bibfield{author}{\bibinfo{person}{Christopher~A. Le~Dantec}.} \bibinfo{year}{2016}\natexlab{}.
\newblock \showarticletitle{Design through collective action collective action through design}.
\newblock \bibinfo{journal}{\emph{Interactions}} \bibinfo{volume}{24}, \bibinfo{number}{1} (\bibinfo{date}{Dec.} \bibinfo{year}{2016}), \bibinfo{pages}{24–30}.
\newblock
\showISSN{1072-5520}
\urldef\tempurl%
\url{https://doi.org/10.1145/3018005}
\showDOI{\tempurl}


\bibitem[Le~Dantec and Fox(2015)]%
        {LeDantec_Fox_2015}
\bibfield{author}{\bibinfo{person}{Christopher~A. Le~Dantec} {and} \bibinfo{person}{Sarah Fox}.} \bibinfo{year}{2015}\natexlab{}.
\newblock \showarticletitle{Strangers at the Gate: Gaining Access, Building Rapport, and Co-Constructing Community-Based Research}. In \bibinfo{booktitle}{\emph{Proceedings of the 18th ACM Conference on Computer Supported Cooperative Work \& Social Computing}} \emph{(\bibinfo{series}{CSCW ’15})}. \bibinfo{publisher}{Association for Computing Machinery}, \bibinfo{address}{New York, NY, USA}, \bibinfo{pages}{1348–1358}.
\newblock
\showISBNx{978-1-4503-2922-4}
\urldef\tempurl%
\url{https://doi.org/10.1145/2675133.2675147}
\showDOI{\tempurl}


\bibitem[Lean(2021)]%
        {Lean_2021}
\bibfield{author}{\bibinfo{person}{Marion Lean}.} \bibinfo{year}{2021}\natexlab{}.
\newblock \showarticletitle{Materialising Data Feminism – How Textile Designers Are Using Materials to Explore Data Experience}.
\newblock \bibinfo{journal}{\emph{Journal of Textile Design Research and Practice}} \bibinfo{volume}{9}, \bibinfo{number}{2} (\bibinfo{date}{May} \bibinfo{year}{2021}), \bibinfo{pages}{184–209}.
\newblock
\showISSN{2051-1787}
\urldef\tempurl%
\url{https://doi.org/10.1080/20511787.2021.1928987}
\showDOI{\tempurl}


\bibitem[Leavy et~al\mbox{.}(2021)]%
        {Leavy_2021}
\bibfield{author}{\bibinfo{person}{Susan Leavy}, \bibinfo{person}{Eugenia Siapera}, {and} \bibinfo{person}{Barry O’Sullivan}.} \bibinfo{year}{2021}\natexlab{}.
\newblock \showarticletitle{Ethical Data Curation for AI: An Approach based on Feminist Epistemology and Critical Theories of Race}. In \bibinfo{booktitle}{\emph{Proceedings of the 2021 AAAI/ACM Conference on AI, Ethics, and Society}} \emph{(\bibinfo{series}{AIES ’21})}. \bibinfo{publisher}{Association for Computing Machinery}, \bibinfo{address}{New York, NY, USA}, \bibinfo{pages}{695–703}.
\newblock
\showISBNx{978-1-4503-8473-5}
\urldef\tempurl%
\url{https://doi.org/10.1145/3461702.3462598}
\showDOI{\tempurl}


\bibitem[Lee et~al\mbox{.}(2019)]%
        {Lee_2019}
\bibfield{author}{\bibinfo{person}{Min~Kyung Lee}, \bibinfo{person}{Daniel Kusbit}, \bibinfo{person}{Anson Kahng}, \bibinfo{person}{Ji~Tae Kim}, \bibinfo{person}{Xinran Yuan}, \bibinfo{person}{Allissa Chan}, \bibinfo{person}{Daniel See}, \bibinfo{person}{Ritesh Noothigattu}, \bibinfo{person}{Siheon Lee}, \bibinfo{person}{Alexandros Psomas}, {and} \bibinfo{person}{Ariel~D. Procaccia}.} \bibinfo{year}{2019}\natexlab{}.
\newblock \showarticletitle{WeBuildAI: Participatory Framework for Algorithmic Governance}.
\newblock \bibinfo{journal}{\emph{Proceedings of the ACM on Human-Computer Interaction}} \bibinfo{volume}{3}, \bibinfo{number}{CSCW} (\bibinfo{date}{Nov.} \bibinfo{year}{2019}), \bibinfo{pages}{181:1--181:35}.
\newblock
\urldef\tempurl%
\url{https://doi.org/10.1145/3359283}
\showDOI{\tempurl}


\bibitem[Li et~al\mbox{.}(2018)]%
        {li2018working}
\bibfield{author}{\bibinfo{person}{Hanlin Li}, \bibinfo{person}{Lynn Dombrowski}, {and} \bibinfo{person}{Erin Brady}.} \bibinfo{year}{2018}\natexlab{}.
\newblock \showarticletitle{Working toward empowering a community: How immigrant-focused nonprofit organizations use Twitter during political conflicts}. In \bibinfo{booktitle}{\emph{Proceedings of the 2018 ACM International Conference on Supporting Group Work}}. \bibinfo{pages}{335--346}.
\newblock
\urldef\tempurl%
\url{https://doi.org/10.1145/3148330.3148336}
\showDOI{\tempurl}


\bibitem[Li et~al\mbox{.}(2021)]%
        {li_ai_2021}
\bibfield{author}{\bibinfo{person}{Victor O.~K. Li}, \bibinfo{person}{Jacqueline C.~K. Lam}, {and} \bibinfo{person}{Jiahuan Cui}.} \bibinfo{year}{2021}\natexlab{}.
\newblock \showarticletitle{{AI} for Social Good: {AI} and Big Data Approaches for Environmental Decision-Making}.
\newblock \bibinfo{journal}{\emph{Environmental Science \& Policy}}  \bibinfo{volume}{125} (\bibinfo{date}{11} \bibinfo{year}{2021}), \bibinfo{pages}{241--246}.
\newblock
\showISSN{1462-9011}
\urldef\tempurl%
\url{https://doi.org/10.1016/j.envsci.2021.09.001}
\showDOI{\tempurl}


\bibitem[Liang et~al\mbox{.}(2023)]%
        {Liang_2023}
\bibfield{author}{\bibinfo{person}{Calvin~Alan Liang}, \bibinfo{person}{Emily Tseng}, \bibinfo{person}{Akeiylah Dewitt}, \bibinfo{person}{Yasmine Kotturi}, \bibinfo{person}{Sucheta Ghoshal}, \bibinfo{person}{Angela D.~R. Smith}, \bibinfo{person}{Marisol Wong-Villacres}, \bibinfo{person}{Lauren Wilcox}, {and} \bibinfo{person}{Sheena Erete}.} \bibinfo{year}{2023}\natexlab{}.
\newblock \showarticletitle{Surfacing Structural Barriers to Community-Collaborative Approaches in Human-Computer Interaction}. In \bibinfo{booktitle}{\emph{Companion Publication of the 2023 Conference on Computer Supported Cooperative Work and Social Computing}} \emph{(\bibinfo{series}{CSCW ’23 Companion})}. \bibinfo{publisher}{Association for Computing Machinery}, \bibinfo{address}{New York, NY, USA}, \bibinfo{pages}{542–546}.
\newblock
\showISBNx{9798400701290}
\urldef\tempurl%
\url{https://doi.org/10.1145/3584931.3611294}
\showDOI{\tempurl}


\bibitem[Lin and Jackson(2023)]%
        {Lin_Jackson_2023}
\bibfield{author}{\bibinfo{person}{Cindy~Kaiying Lin} {and} \bibinfo{person}{Steven~J. Jackson}.} \bibinfo{year}{2023}\natexlab{}.
\newblock \showarticletitle{From Bias to Repair: Error as a Site of Collaboration and Negotiation in Applied Data Science Work}.
\newblock \bibinfo{journal}{\emph{Proceedings of the ACM on Human-Computer Interaction}} \bibinfo{volume}{7}, \bibinfo{number}{CSCW1} (\bibinfo{date}{April} \bibinfo{year}{2023}), \bibinfo{pages}{131:1--131:32}.
\newblock
\urldef\tempurl%
\url{https://doi.org/10.1145/3579607}
\showDOI{\tempurl}


\bibitem[Lodato and DiSalvo(2018)]%
        {Lodato2018institutional}
\bibfield{author}{\bibinfo{person}{Thomas Lodato} {and} \bibinfo{person}{Carl DiSalvo}.} \bibinfo{year}{2018}\natexlab{}.
\newblock \showarticletitle{Institutional constraints: the forms and limits of participatory design in the public realm}. In \bibinfo{booktitle}{\emph{Proceedings of the 15th Participatory Design Conference: Full Papers - Volume 1}} \emph{(\bibinfo{series}{PDC ’18})}. \bibinfo{publisher}{Association for Computing Machinery}, \bibinfo{address}{New York, NY, USA}, \bibinfo{pages}{1–12}.
\newblock
\showISBNx{978-1-4503-6371-6}
\urldef\tempurl%
\url{https://doi.org/10.1145/3210586.3210595}
\showDOI{\tempurl}


\bibitem[Lupetti and Murray-Rust(2024)]%
        {Lupetti2024unmaking}
\bibfield{author}{\bibinfo{person}{Maria~Luce Lupetti} {and} \bibinfo{person}{Dave Murray-Rust}.} \bibinfo{year}{2024}\natexlab{}.
\newblock \showarticletitle{(Un)making AI Magic: a Design Taxonomy}.
\newblock  (\bibinfo{date}{March} \bibinfo{year}{2024}).
\newblock
\urldef\tempurl%
\url{https://doi.org/10.1145/3613904.3641954}
\showDOI{\tempurl}
\newblock
\shownote{arXiv:2403.15216 [cs]}.


\bibitem[Magalhães and Couldry(2021)]%
        {magalhaes_giving_2021}
\bibfield{author}{\bibinfo{person}{João~Carlos Magalhães} {and} \bibinfo{person}{Nick Couldry}.} \bibinfo{year}{2021}\natexlab{}.
\newblock \showarticletitle{Giving by Taking Away: Big Tech, Data Colonialism, and the Reconfiguration of Social Good}.
\newblock \bibinfo{journal}{\emph{International Journal of Communication}} \bibinfo{volume}{15}, \bibinfo{number}{0} (\bibinfo{date}{1} \bibinfo{year}{2021}), \bibinfo{pages}{20}.
\newblock
\showISSN{1932-8036}
\urldef\tempurl%
\url{https://ijoc.org/index.php/ijoc/article/view/15995}
\showURL{%
\tempurl}
\newblock
\shownote{Number: 0}.


\bibitem[Maxwell et~al\mbox{.}(2016)]%
        {maxwell_data_2016}
\bibfield{author}{\bibinfo{person}{Nan~L. Maxwell}, \bibinfo{person}{Dana Rotz}, {and} \bibinfo{person}{Christina Garcia}.} \bibinfo{year}{2016}\natexlab{}.
\newblock \showarticletitle{Data and Decision Making: Same Organization, Different Perceptions; Different Organizations, Different Perceptions}.
\newblock \bibinfo{journal}{\emph{American Journal of Evaluation}} \bibinfo{volume}{37}, \bibinfo{number}{4} (\bibinfo{date}{Dec.} \bibinfo{year}{2016}), \bibinfo{pages}{463--485}.
\newblock
\showISSN{1098-2140}
\urldef\tempurl%
\url{https://doi.org/10.1177/1098214015623634}
\showDOI{\tempurl}
\newblock
\shownote{Publisher: {SAGE} Publications Inc}.


\bibitem[Meng and DiSalvo(2018)]%
        {meng2018grassroots}
\bibfield{author}{\bibinfo{person}{Amanda Meng} {and} \bibinfo{person}{Carl DiSalvo}.} \bibinfo{year}{2018}\natexlab{}.
\newblock \showarticletitle{Grassroots resource mobilization through counter-data action}.
\newblock \bibinfo{journal}{\emph{Big Data \& Society}} \bibinfo{volume}{5}, \bibinfo{number}{2} (\bibinfo{year}{2018}).
\newblock
\urldef\tempurl%
\url{https://doi.org/10.1177/2053951718796862}
\showDOI{\tempurl}


\bibitem[Meng et~al\mbox{.}(2019)]%
        {meng2019collaborative}
\bibfield{author}{\bibinfo{person}{Amanda Meng}, \bibinfo{person}{Carl DiSalvo}, {and} \bibinfo{person}{Ellen Zegura}.} \bibinfo{year}{2019}\natexlab{}.
\newblock \showarticletitle{Collaborative data work towards a caring democracy}.
\newblock \bibinfo{journal}{\emph{Proceedings of the ACM on Human-Computer Interaction}} \bibinfo{volume}{3}, \bibinfo{number}{CSCW} (\bibinfo{year}{2019}), \bibinfo{pages}{1--23}.
\newblock
\urldef\tempurl%
\url{https://doi.org/10.1145/3359144}
\showDOI{\tempurl}


\bibitem[Miceli and Posada(2022)]%
        {miceli2022data}
\bibfield{author}{\bibinfo{person}{Milagros Miceli} {and} \bibinfo{person}{Julian Posada}.} \bibinfo{year}{2022}\natexlab{}.
\newblock \showarticletitle{The Data-Production Dispositif}.
\newblock \bibinfo{journal}{\emph{Proceedings of the ACM on Human-Computer Interaction}} \bibinfo{volume}{6}, \bibinfo{number}{CSCW2} (\bibinfo{year}{2022}), \bibinfo{pages}{1--37}.
\newblock
\urldef\tempurl%
\url{https://doi.org/10.1145/3555561}
\showDOI{\tempurl}


\bibitem[Miceli et~al\mbox{.}(2022)]%
        {miceli2022studying}
\bibfield{author}{\bibinfo{person}{Milagros Miceli}, \bibinfo{person}{Julian Posada}, {and} \bibinfo{person}{Tianling Yang}.} \bibinfo{year}{2022}\natexlab{}.
\newblock \showarticletitle{Studying up machine learning data: Why talk about bias when we mean power?}
\newblock \bibinfo{journal}{\emph{Proceedings of the ACM on Human-Computer Interaction}} \bibinfo{volume}{6}, \bibinfo{number}{GROUP} (\bibinfo{year}{2022}), \bibinfo{pages}{1--14}.
\newblock
\urldef\tempurl%
\url{https://doi.org/10.1145/3492853}
\showDOI{\tempurl}


\bibitem[Miceli et~al\mbox{.}(2020)]%
        {miceli2020between}
\bibfield{author}{\bibinfo{person}{Milagros Miceli}, \bibinfo{person}{Martin Schuessler}, {and} \bibinfo{person}{Tianling Yang}.} \bibinfo{year}{2020}\natexlab{}.
\newblock \showarticletitle{Between subjectivity and imposition: Power dynamics in data annotation for computer vision}.
\newblock \bibinfo{journal}{\emph{Proceedings of the ACM on Human-Computer Interaction}} \bibinfo{volume}{4}, \bibinfo{number}{CSCW2} (\bibinfo{year}{2020}), \bibinfo{pages}{1--25}.
\newblock
\urldef\tempurl%
\url{https://doi.org/10.1145/3415186}
\showDOI{\tempurl}


\bibitem[Muhammad et~al\mbox{.}(2015)]%
        {Muhammad2015reflections}
\bibfield{author}{\bibinfo{person}{Michael Muhammad}, \bibinfo{person}{Nina Wallerstein}, \bibinfo{person}{Andrew~L. Sussman}, \bibinfo{person}{Magdalena Avila}, \bibinfo{person}{Lorenda Belone}, {and} \bibinfo{person}{Bonnie Duran}.} \bibinfo{year}{2015}\natexlab{}.
\newblock \showarticletitle{Reflections on Researcher Identity and Power: The Impact of Positionality on Community Based Participatory Research (CBPR) Processes and Outcomes}.
\newblock \bibinfo{journal}{\emph{Critical sociology}} \bibinfo{volume}{41}, \bibinfo{number}{7–8} (\bibinfo{date}{Nov.} \bibinfo{year}{2015}), \bibinfo{pages}{1045–1063}.
\newblock
\showISSN{0896-9205}
\urldef\tempurl%
\url{https://doi.org/10.1177/0896920513516025}
\showDOI{\tempurl}


\bibitem[Naderifar et~al\mbox{.}(2017)]%
        {Naderifar_Goli_Ghaljaie_2017}
\bibfield{author}{\bibinfo{person}{Mahin Naderifar}, \bibinfo{person}{Hamideh Goli}, {and} \bibinfo{person}{Fereshteh Ghaljaie}.} \bibinfo{year}{2017}\natexlab{}.
\newblock \showarticletitle{Snowball Sampling: A Purposeful Method of Sampling in Qualitative Research}.
\newblock \bibinfo{journal}{\emph{Strides in Development of Medical Education}} \bibinfo{volume}{14}, \bibinfo{number}{3} (\bibinfo{date}{Sept.} \bibinfo{year}{2017}).
\newblock
\showISSN{2645-3525}
\urldef\tempurl%
\url{https://doi.org/10.5812/sdme.67670}
\showDOI{\tempurl}


\bibitem[Okolo et~al\mbox{.}(2024)]%
        {Okolo_2024}
\bibfield{author}{\bibinfo{person}{Chinasa~T. Okolo}, \bibinfo{person}{Dhruv Agarwal}, \bibinfo{person}{Nicola Dell}, {and} \bibinfo{person}{Aditya Vashistha}.} \bibinfo{year}{2024}\natexlab{}.
\newblock \showarticletitle{“If it is easy to understand then it will have value”: Examining Perceptions of Explainable AI with Community Health Workers in Rural India}.
\newblock \bibinfo{journal}{\emph{Proceedings of the ACM on Human-Computer Interaction}} \bibinfo{volume}{8}, \bibinfo{number}{CSCW1} (\bibinfo{date}{April} \bibinfo{year}{2024}), \bibinfo{pages}{71:1--71:28}.
\newblock
\urldef\tempurl%
\url{https://doi.org/10.1145/3637348}
\showDOI{\tempurl}


\bibitem[Okolo et~al\mbox{.}(2021)]%
        {okolo_it_2021}
\bibfield{author}{\bibinfo{person}{Chinasa~T. Okolo}, \bibinfo{person}{Srujana Kamath}, \bibinfo{person}{Nicola Dell}, {and} \bibinfo{person}{Aditya Vashistha}.} \bibinfo{year}{2021}\natexlab{}.
\newblock \showarticletitle{“It cannot do all of my work”: Community Health Worker Perceptions of {AI}-Enabled Mobile Health Applications in Rural India}. In \bibinfo{booktitle}{\emph{Proceedings of the 2021 {CHI} Conference on Human Factors in Computing Systems}} (New York, {NY}, {USA}, 2021-05-07) \emph{(\bibinfo{series}{{CHI} '21})}. \bibinfo{publisher}{Association for Computing Machinery}, \bibinfo{pages}{1--20}.
\newblock
\showISBNx{978-1-4503-8096-6}
\urldef\tempurl%
\url{https://doi.org/10.1145/3411764.3445420}
\showDOI{\tempurl}


\bibitem[Okolo and Lin(2024)]%
        {Okolo2024you}
\bibfield{author}{\bibinfo{person}{Chinasa~T. Okolo} {and} \bibinfo{person}{Hongjin Lin}.} \bibinfo{year}{2024}\natexlab{}.
\newblock \showarticletitle{“You can’t build what you don’t understand”: Practitioner Perspectives on Explainable AI in the Global South}. In \bibinfo{booktitle}{\emph{Extended Abstracts of the 2024 CHI Conference on Human Factors in Computing Systems}} \emph{(\bibinfo{series}{CHI EA ’24})}. \bibinfo{publisher}{Association for Computing Machinery}, \bibinfo{address}{New York, NY, USA}, \bibinfo{pages}{1–10}.
\newblock
\showISBNx{9798400703317}
\urldef\tempurl%
\url{https://doi.org/10.1145/3613905.3651080}
\showDOI{\tempurl}


\bibitem[Pandit(1996)]%
        {Pandit_1996}
\bibfield{author}{\bibinfo{person}{Naresh Pandit}.} \bibinfo{year}{1996}\natexlab{}.
\newblock \showarticletitle{The Creation of Theory: A Recent Application of the Grounded Theory Method}.
\newblock \bibinfo{journal}{\emph{The Qualitative Report}} \bibinfo{volume}{2}, \bibinfo{number}{4} (\bibinfo{date}{Dec.} \bibinfo{year}{1996}), \bibinfo{pages}{1–15}.
\newblock
\showISSN{1052-0147}
\urldef\tempurl%
\url{https://doi.org/10.46743/2160-3715/1996.2054}
\showDOI{\tempurl}


\bibitem[Passi and Sengers(2020)]%
        {passi2020making}
\bibfield{author}{\bibinfo{person}{Samir Passi} {and} \bibinfo{person}{Phoebe Sengers}.} \bibinfo{year}{2020}\natexlab{}.
\newblock \showarticletitle{Making data science systems work}.
\newblock \bibinfo{journal}{\emph{Big Data \& Society}} \bibinfo{volume}{7}, \bibinfo{number}{2} (\bibinfo{year}{2020}).
\newblock
\urldef\tempurl%
\url{https://doi.org/10.1177/2053951720939605}
\showDOI{\tempurl}


\bibitem[Paudel and Soden(2023)]%
        {Paudel_Soden_2023}
\bibfield{author}{\bibinfo{person}{Shreyasha Paudel} {and} \bibinfo{person}{Robert Soden}.} \bibinfo{year}{2023}\natexlab{}.
\newblock \showarticletitle{Reimagining Open Data during Disaster Response: Applying a Feminist Lens to Three Open Data Projects in Post-Earthquake Nepal}.
\newblock \bibinfo{journal}{\emph{Proceedings of the ACM on Human-Computer Interaction}} \bibinfo{volume}{7}, \bibinfo{number}{CSCW1} (\bibinfo{date}{April} \bibinfo{year}{2023}), \bibinfo{pages}{86:1--86:25}.
\newblock
\urldef\tempurl%
\url{https://doi.org/10.1145/3579519}
\showDOI{\tempurl}


\bibitem[Perrault et~al\mbox{.}(2020)]%
        {perrault_ai_2020}
\bibfield{author}{\bibinfo{person}{Andrew Perrault}, \bibinfo{person}{Fei Fang}, \bibinfo{person}{Arunesh Sinha}, {and} \bibinfo{person}{Milind Tambe}.} \bibinfo{year}{2020}\natexlab{}.
\newblock \showarticletitle{{AI} for Social Impact: Learning and Planning in the Data-to-Deployment Pipeline}.
\newblock \bibinfo{journal}{\emph{{AI} Magazine}} \bibinfo{volume}{41}, \bibinfo{number}{4} (\bibinfo{date}{Dec.} \bibinfo{year}{2020}), \bibinfo{pages}{3--16}.
\newblock
\showISSN{0738-4602, 2371-9621}
\urldef\tempurl%
\url{https://doi.org/10.1609/aimag.v41i4.5296}
\showDOI{\tempurl}
\showeprint[arxiv]{2001.00088 [cs]}


\bibitem[Pruss(2023)]%
        {pruss_ghosting_2023}
\bibfield{author}{\bibinfo{person}{Dasha Pruss}.} \bibinfo{year}{2023}\natexlab{}.
\newblock \showarticletitle{Ghosting the Machine: Judicial Resistance to a Recidivism Risk Assessment Instrument}. In \bibinfo{booktitle}{\emph{Proceedings of the 2023 {ACM} Conference on Fairness, Accountability, and Transparency}} (New York, {NY}, {USA}, 2023-06-12) \emph{(\bibinfo{series}{{FAccT} '23})}. \bibinfo{publisher}{Association for Computing Machinery}, \bibinfo{pages}{312--323}.
\newblock
\showISBNx{9798400701924}
\urldef\tempurl%
\url{https://doi.org/10.1145/3593013.3593999}
\showDOI{\tempurl}


\bibitem[Racadio et~al\mbox{.}(2014)]%
        {Racadio_2014}
\bibfield{author}{\bibinfo{person}{Robert Racadio}, \bibinfo{person}{Emma~J. Rose}, {and} \bibinfo{person}{Beth~E. Kolko}.} \bibinfo{year}{2014}\natexlab{}.
\newblock \showarticletitle{Research at the margin: participatory design and community based participatory research} \emph{(\bibinfo{series}{PDC ’14})}. \bibinfo{address}{New York, NY, USA}, \bibinfo{pages}{49–52}.
\newblock
\showISBNx{978-1-4503-3214-9}
\urldef\tempurl%
\url{https://doi.org/10.1145/2662155.2662188}
\showDOI{\tempurl}


\bibitem[Radhakrishnan(2021)]%
        {radhakrishnan_experiments_2021}
\bibfield{author}{\bibinfo{person}{Radhika Radhakrishnan}.} \bibinfo{year}{2021}\natexlab{}.
\newblock \showarticletitle{Experiments with Social Good: Feminist Critiques of Artificial Intelligence in Healthcare in India}.
\newblock \bibinfo{journal}{\emph{Catalyst: Feminism, Theory, Technoscience}} \bibinfo{volume}{7}, \bibinfo{number}{2} (\bibinfo{date}{Oct.} \bibinfo{year}{2021}).
\newblock
\showISSN{2380-3312}
\urldef\tempurl%
\url{https://doi.org/10.28968/cftt.v7i2.34916}
\showDOI{\tempurl}
\newblock
\shownote{Number: 2}.


\bibitem[Ramesh et~al\mbox{.}(2022)]%
        {ramesh2022platform}
\bibfield{author}{\bibinfo{person}{Divya Ramesh}, \bibinfo{person}{Vaishnav Kameswaran}, \bibinfo{person}{Ding Wang}, {and} \bibinfo{person}{Nithya Sambasivan}.} \bibinfo{year}{2022}\natexlab{}.
\newblock \showarticletitle{How platform-user power relations shape algorithmic accountability: A case study of instant loan platforms and financially stressed users in India}. In \bibinfo{booktitle}{\emph{Proceedings of the 2022 ACM Conference on Fairness, Accountability, and Transparency}}. \bibinfo{pages}{1917--1928}.
\newblock
\urldef\tempurl%
\url{https://doi.org/10.1145/3531146.3533237}
\showDOI{\tempurl}


\bibitem[Saha et~al\mbox{.}(2022a)]%
        {saha2022commissioning}
\bibfield{author}{\bibinfo{person}{Manika Saha}, \bibinfo{person}{Tom Bartindale}, \bibinfo{person}{Sharifa Sultana}, \bibinfo{person}{Gillian Oliver}, \bibinfo{person}{Dan Richardson}, \bibinfo{person}{Shakuntala~Haraksingh Thilsted}, \bibinfo{person}{Syed~Ishtiaque Ahmed}, {and} \bibinfo{person}{Patrick Olivier}.} \bibinfo{year}{2022}\natexlab{a}.
\newblock \showarticletitle{Commissioning Development: Grantmaking, Community Voices, and their Implications for ICTD}. In \bibinfo{booktitle}{\emph{Proceedings of the 2022 International Conference on Information and Communication Technologies and Development}}. \bibinfo{pages}{1--18}.
\newblock
\urldef\tempurl%
\url{https://doi.org/10.1145/3572334.3572402}
\showDOI{\tempurl}


\bibitem[Saha et~al\mbox{.}(2022b)]%
        {Saha2022towards}
\bibfield{author}{\bibinfo{person}{Manika Saha}, \bibinfo{person}{Delvin Varghese}, \bibinfo{person}{Tom Bartindale}, \bibinfo{person}{Shakuntala~Haraksingh Thilsted}, \bibinfo{person}{Syed~Ishtiaque Ahmed}, {and} \bibinfo{person}{Patrick Olivier}.} \bibinfo{year}{2022}\natexlab{b}.
\newblock \showarticletitle{Towards Sustainable ICTD in Bangladesh: Understanding the Program and Policy Landscape and Its Implications for CSCW and HCI}.
\newblock \bibinfo{journal}{\emph{Proc. ACM Hum.-Comput. Interact.}} \bibinfo{volume}{6}, \bibinfo{number}{CSCW1} (\bibinfo{date}{April} \bibinfo{year}{2022}), \bibinfo{pages}{126:1--126:31}.
\newblock
\urldef\tempurl%
\url{https://doi.org/10.1145/3512973}
\showDOI{\tempurl}


\bibitem[Sambasivan et~al\mbox{.}(2021)]%
        {sambasivan_everyone_2021}
\bibfield{author}{\bibinfo{person}{Nithya Sambasivan}, \bibinfo{person}{Shivani Kapania}, \bibinfo{person}{Hannah Highfill}, \bibinfo{person}{Diana Akrong}, \bibinfo{person}{Praveen Paritosh}, {and} \bibinfo{person}{Lora~M Aroyo}.} \bibinfo{year}{2021}\natexlab{}.
\newblock \showarticletitle{“Everyone wants to do the model work, not the data work”: Data Cascades in High-Stakes {AI}}. In \bibinfo{booktitle}{\emph{Proceedings of the 2021 {CHI} Conference on Human Factors in Computing Systems}} (New York, {NY}, {USA}, 2021-05-07) \emph{(\bibinfo{series}{{CHI} '21})}. \bibinfo{publisher}{Association for Computing Machinery}, \bibinfo{pages}{1--15}.
\newblock
\showISBNx{978-1-4503-8096-6}
\urldef\tempurl%
\url{https://doi.org/10.1145/3411764.3445518}
\showDOI{\tempurl}


\bibitem[Sambasivan and Veeraraghavan(2022)]%
        {sambasivan_deskilling_2022}
\bibfield{author}{\bibinfo{person}{Nithya Sambasivan} {and} \bibinfo{person}{Rajesh Veeraraghavan}.} \bibinfo{year}{2022}\natexlab{}.
\newblock \showarticletitle{The Deskilling of Domain Expertise in {AI} Development}. In \bibinfo{booktitle}{\emph{Proceedings of the 2022 {CHI} Conference on Human Factors in Computing Systems}} (New York, {NY}, {USA}, 2022-04-29) \emph{(\bibinfo{series}{{CHI} '22})}. \bibinfo{publisher}{Association for Computing Machinery}, \bibinfo{pages}{1--14}.
\newblock
\showISBNx{978-1-4503-9157-3}
\urldef\tempurl%
\url{https://doi.org/10.1145/3491102.3517578}
\showDOI{\tempurl}


\bibitem[Sandberg et~al\mbox{.}(2023)]%
        {sandberg2023re}
\bibfield{author}{\bibinfo{person}{Billie Sandberg}, \bibinfo{person}{Laura~C Hand}, {and} \bibinfo{person}{Andrew Russo}.} \bibinfo{year}{2023}\natexlab{}.
\newblock \showarticletitle{Re-envisioning the role of “big data” in the nonprofit sector: A data feminist perspective}.
\newblock \bibinfo{journal}{\emph{VOLUNTAS: International Journal of Voluntary and Nonprofit Organizations}} \bibinfo{volume}{34}, \bibinfo{number}{5} (\bibinfo{year}{2023}), \bibinfo{pages}{1094--1105}.
\newblock
\urldef\tempurl%
\url{https://doi.org/10.1007/s11266-022-00529-9}
\showDOI{\tempurl}


\bibitem[Schelenz and Pawelec(2022)]%
        {schelenz_information_2022}
\bibfield{author}{\bibinfo{person}{Laura Schelenz} {and} \bibinfo{person}{Maria Pawelec}.} \bibinfo{year}{2022}\natexlab{}.
\newblock \showarticletitle{Information and Communication Technologies for Development ({ICT}4D) critique}.
\newblock \bibinfo{journal}{\emph{Information Technology for Development}} \bibinfo{volume}{28}, \bibinfo{number}{1} (\bibinfo{year}{2022}), \bibinfo{pages}{165--188}.
\newblock
\showISSN{0268-1102}
\urldef\tempurl%
\url{https://doi.org/10.1080/02681102.2021.1937473}
\showDOI{\tempurl}


\bibitem[Shi et~al\mbox{.}(2020)]%
        {shi_artificial_2020}
\bibfield{author}{\bibinfo{person}{Zheyuan~Ryan Shi}, \bibinfo{person}{Claire Wang}, {and} \bibinfo{person}{Fei Fang}.} \bibinfo{year}{2020}\natexlab{}.
\newblock \showarticletitle{Artificial Intelligence for Social Good: A Survey}.
\newblock \bibinfo{journal}{\emph{{ArXiv}}} (\bibinfo{date}{Jan.} \bibinfo{year}{2020}).
\newblock
\urldef\tempurl%
\url{https://arxiv.org/abs/2001.01818}
\showURL{%
\tempurl}


\bibitem[Sims(2017)]%
        {Sims_2017}
\bibfield{author}{\bibinfo{person}{Christo Sims}.} \bibinfo{year}{2017}\natexlab{}.
\newblock \bibinfo{booktitle}{\emph{Disruptive Fixation: School Reform and the Pitfalls of Techno-Idealism}}.
\newblock \bibinfo{publisher}{Princeton University Press}.
\newblock
\showISBNx{978-0-691-16399-4}


\bibitem[Sum et~al\mbox{.}(2023)]%
        {Sum_2023}
\bibfield{author}{\bibinfo{person}{Cella~M Sum}, \bibinfo{person}{Anh-Ton Tran}, \bibinfo{person}{Jessica Lin}, \bibinfo{person}{Rachel Kuo}, \bibinfo{person}{Cynthia~L Bennett}, \bibinfo{person}{Christina Harrington}, {and} \bibinfo{person}{Sarah~E Fox}.} \bibinfo{year}{2023}\natexlab{}.
\newblock \showarticletitle{Translation as (Re)mediation: How Ethnic Community-Based Organizations Negotiate Legitimacy}. In \bibinfo{booktitle}{\emph{Proceedings of the 2023 CHI Conference on Human Factors in Computing Systems}}. \bibinfo{publisher}{ACM}, \bibinfo{address}{Hamburg Germany}, \bibinfo{pages}{1–14}.
\newblock
\showISBNx{978-1-4503-9421-5}
\urldef\tempurl%
\url{https://doi.org/10.1145/3544548.3581280}
\showDOI{\tempurl}


\bibitem[Sun et~al\mbox{.}(2023)]%
        {sun2023care}
\bibfield{author}{\bibinfo{person}{Yuling Sun}, \bibinfo{person}{Xiaojuan Ma}, \bibinfo{person}{Silvia Lindtner}, {and} \bibinfo{person}{Liang He}.} \bibinfo{year}{2023}\natexlab{}.
\newblock \showarticletitle{Care Workers' Wellbeing in Data-Driven Healthcare Workplace: Identity, Agency, and Social Justice}.
\newblock \bibinfo{journal}{\emph{Proceedings of the ACM on Human-Computer Interaction}} \bibinfo{volume}{7}, \bibinfo{number}{CSCW2} (\bibinfo{year}{2023}), \bibinfo{pages}{1--29}.
\newblock
\urldef\tempurl%
\url{https://doi.org/10.1145/3610178}
\showDOI{\tempurl}


\bibitem[Suresh et~al\mbox{.}(2022)]%
        {Suresh_2022}
\bibfield{author}{\bibinfo{person}{Harini Suresh}, \bibinfo{person}{Rajiv Movva}, \bibinfo{person}{Amelia~Lee Dogan}, \bibinfo{person}{Rahul Bhargava}, \bibinfo{person}{Isadora Cruxen}, \bibinfo{person}{Angeles~Martinez Cuba}, \bibinfo{person}{Guilia Taurino}, \bibinfo{person}{Wonyoung So}, {and} \bibinfo{person}{Catherine D’Ignazio}.} \bibinfo{year}{2022}\natexlab{}.
\newblock \showarticletitle{Towards Intersectional Feminist and Participatory ML: A Case Study in Supporting Feminicide Counterdata Collection}. In \bibinfo{booktitle}{\emph{Proceedings of the 2022 ACM Conference on Fairness, Accountability, and Transparency}} \emph{(\bibinfo{series}{FAccT ’22})}. \bibinfo{publisher}{Association for Computing Machinery}, \bibinfo{address}{New York, NY, USA}, \bibinfo{pages}{667–678}.
\newblock
\showISBNx{978-1-4503-9352-2}
\urldef\tempurl%
\url{https://doi.org/10.1145/3531146.3533132}
\showDOI{\tempurl}


\bibitem[Susha et~al\mbox{.}(2019)]%
        {susha_data_2019}
\bibfield{author}{\bibinfo{person}{Iryna Susha}, \bibinfo{person}{Åke Grönlund}, {and} \bibinfo{person}{Rob Van~Tulder}.} \bibinfo{year}{2019}\natexlab{}.
\newblock \showarticletitle{Data driven social partnerships: Exploring an emergent trend in search of research challenges and questions}.
\newblock \bibinfo{journal}{\emph{Government Information Quarterly}} \bibinfo{volume}{36}, \bibinfo{number}{1} (\bibinfo{date}{Jan.} \bibinfo{year}{2019}), \bibinfo{pages}{112--128}.
\newblock
\showISSN{0740-624X}
\urldef\tempurl%
\url{https://doi.org/10.1016/j.giq.2018.11.002}
\showDOI{\tempurl}


\bibitem[Susha et~al\mbox{.}(2023)]%
        {susha_achieving_2023}
\bibfield{author}{\bibinfo{person}{Iryna Susha}, \bibinfo{person}{Boriana Rukanova}, \bibinfo{person}{Anneke Zuiderwijk}, \bibinfo{person}{J.~Ramon Gil-Garcia}, {and} \bibinfo{person}{Mila Gasco~Hernandez}.} \bibinfo{year}{2023}\natexlab{}.
\newblock \showarticletitle{Achieving voluntary data sharing in cross sector partnerships: Three partnership models}.
\newblock \bibinfo{journal}{\emph{Information and Organization}} \bibinfo{volume}{33}, \bibinfo{number}{1} (\bibinfo{date}{March} \bibinfo{year}{2023}), \bibinfo{pages}{100448}.
\newblock
\showISSN{1471-7727}
\urldef\tempurl%
\url{https://doi.org/10.1016/j.infoandorg.2023.100448}
\showDOI{\tempurl}


\bibitem[Tanweer and Fiore-Gartland(2017)]%
        {tanweer_cross-sector_2017}
\bibfield{author}{\bibinfo{person}{Anissa Tanweer} {and} \bibinfo{person}{Brittany Fiore-Gartland}.} \bibinfo{year}{2017}\natexlab{}.
\newblock \bibinfo{title}{Cross-sector Collaboration in Data Science for Social Good: Opportunities, Challenges, and Open Questions Raised by Working with Academic Researchers}.
\newblock
\newblock
\urldef\tempurl%
\url{https://www.dssgfellowship.org//wp-content/uploads/2017/09/tanweer.pdf}
\showURL{%
\tempurl}


\bibitem[Thakkar et~al\mbox{.}(2022)]%
        {thakkar2022machine}
\bibfield{author}{\bibinfo{person}{Divy Thakkar}, \bibinfo{person}{Azra Ismail}, \bibinfo{person}{Pratyush Kumar}, \bibinfo{person}{Alex Hanna}, \bibinfo{person}{Nithya Sambasivan}, {and} \bibinfo{person}{Neha Kumar}.} \bibinfo{year}{2022}\natexlab{}.
\newblock \showarticletitle{When is machine learning data good?: Valuing in public health datafication}. In \bibinfo{booktitle}{\emph{Proceedings of the 2022 CHI Conference on Human Factors in Computing Systems}}. \bibinfo{pages}{1--16}.
\newblock
\urldef\tempurl%
\url{https://doi.org/10.1145/3491102.3501868}
\showDOI{\tempurl}


\bibitem[Thinyane et~al\mbox{.}(2018)]%
        {thinyane2018critical}
\bibfield{author}{\bibinfo{person}{Mamello Thinyane}, \bibinfo{person}{Karthik Bhat}, \bibinfo{person}{Lauri Goldkind}, {and} \bibinfo{person}{Vikram~Kamath Cannanure}.} \bibinfo{year}{2018}\natexlab{}.
\newblock \showarticletitle{Critical participatory design: reflections on engagement and empowerment in a case of a community based organization}. In \bibinfo{booktitle}{\emph{Proceedings of the 15th Participatory Design Conference: Full Papers-Volume 1}}. \bibinfo{pages}{1--10}.
\newblock
\urldef\tempurl%
\url{https://doi.org/10.1145/3210586.3210601}
\showDOI{\tempurl}


\bibitem[Till et~al\mbox{.}(2022)]%
        {till2022community}
\bibfield{author}{\bibinfo{person}{Sarina Till}, \bibinfo{person}{Jaydon Farao}, \bibinfo{person}{Toshka~Lauren Coleman}, \bibinfo{person}{Londiwe~Deborah Shandu}, \bibinfo{person}{Nonkululeko Khuzwayo}, \bibinfo{person}{Livhuwani Muthelo}, \bibinfo{person}{Masenyani~Oupa Mbombi}, \bibinfo{person}{Mamare Bopane}, \bibinfo{person}{Molebogeng Motlhatlhedi}, \bibinfo{person}{Gugulethu Mabena}, \bibinfo{person}{Alastair Van~Heerden}, \bibinfo{person}{Tebogo~Maria Mothiba}, \bibinfo{person}{Shane Norris}, \bibinfo{person}{Nervo Verdezoto~Dias}, {and} \bibinfo{person}{Melissa Densmore}.} \bibinfo{year}{2022}\natexlab{}.
\newblock \showarticletitle{Community-based co-design across geographic locations and cultures: methodological lessons from co-design workshops in South Africa}. In \bibinfo{booktitle}{\emph{Proceedings of the Participatory Design Conference 2022-Volume 1}}. \bibinfo{pages}{120--132}.
\newblock
\urldef\tempurl%
\url{https://doi.org/10.1145/3536169.3537786}
\showDOI{\tempurl}


\bibitem[Tomašev et~al\mbox{.}(2020)]%
        {tomasev_ai_2020}
\bibfield{author}{\bibinfo{person}{Nenad Tomašev}, \bibinfo{person}{Julien Cornebise}, \bibinfo{person}{Frank Hutter}, \bibinfo{person}{Shakir Mohamed}, \bibinfo{person}{Angela Picciariello}, \bibinfo{person}{Bec Connelly}, \bibinfo{person}{Danielle C.~M. Belgrave}, \bibinfo{person}{Daphne Ezer}, \bibinfo{person}{Fanny Cachat van~der Haert}, \bibinfo{person}{Frank Mugisha}, \bibinfo{person}{Gerald Abila}, \bibinfo{person}{Hiromi Arai}, \bibinfo{person}{Hisham Almiraat}, \bibinfo{person}{Julia Proskurnia}, \bibinfo{person}{Kyle Snyder}, \bibinfo{person}{Mihoko Otake-Matsuura}, \bibinfo{person}{Mustafa Othman}, \bibinfo{person}{Tobias Glasmachers}, \bibinfo{person}{Wilfried~de Wever}, \bibinfo{person}{Yee~Whye Teh}, \bibinfo{person}{Mohammad~Emtiyaz Khan}, \bibinfo{person}{Ruben~De Winne}, \bibinfo{person}{Tom Schaul}, {and} \bibinfo{person}{Claudia Clopath}.} \bibinfo{year}{2020}\natexlab{}.
\newblock \showarticletitle{{AI} for social good: unlocking the opportunity for positive impact}.
\newblock \bibinfo{journal}{\emph{Nature Communications}} \bibinfo{volume}{11}, \bibinfo{number}{1} (\bibinfo{date}{May} \bibinfo{year}{2020}), \bibinfo{pages}{2468}.
\newblock
\showISSN{2041-1723}
\urldef\tempurl%
\url{https://doi.org/10.1038/s41467-020-15871-z}
\showDOI{\tempurl}


\bibitem[Toyama(2015)]%
        {toyama_geek_2015}
\bibfield{author}{\bibinfo{person}{Kentaro Toyama}.} \bibinfo{year}{2015}\natexlab{}.
\newblock \bibinfo{booktitle}{\emph{Geek Heresy: Rescuing Social Change from the Cult of Technology}}.
\newblock \bibinfo{publisher}{{PublicAffairs}}.
\newblock
\showISBNx{978-1-61039-529-8}


\bibitem[Tran et~al\mbox{.}(2022)]%
        {tran2022careful}
\bibfield{author}{\bibinfo{person}{Anh-Ton Tran}, \bibinfo{person}{Ashley Boone}, \bibinfo{person}{Christopher~A Le~Dantec}, {and} \bibinfo{person}{Carl DiSalvo}.} \bibinfo{year}{2022}\natexlab{}.
\newblock \showarticletitle{Careful Data Tinkering}.
\newblock \bibinfo{journal}{\emph{Proceedings of the ACM on Human-Computer Interaction}} \bibinfo{volume}{6}, \bibinfo{number}{CSCW2} (\bibinfo{year}{2022}), \bibinfo{pages}{1--29}.
\newblock
\urldef\tempurl%
\url{https://doi.org/10.1145/3555532}
\showDOI{\tempurl}


\bibitem[Unertl et~al\mbox{.}(2016)]%
        {Unertl_2016}
\bibfield{author}{\bibinfo{person}{Kim~M. Unertl}, \bibinfo{person}{Chris~L. Schaefbauer}, \bibinfo{person}{Terrance~R. Campbell}, \bibinfo{person}{Charles Senteio}, \bibinfo{person}{Katie~A. Siek}, \bibinfo{person}{Suzanne Bakken}, {and} \bibinfo{person}{Tiffany~C. Veinot}.} \bibinfo{year}{2016}\natexlab{}.
\newblock \showarticletitle{Integrating community-based participatory research and informatics approaches to improve the engagement and health of underserved populations}.
\newblock \bibinfo{journal}{\emph{Journal of the American Medical Informatics Association: JAMIA}} \bibinfo{volume}{23}, \bibinfo{number}{1} (\bibinfo{date}{Jan.} \bibinfo{year}{2016}), \bibinfo{pages}{60–73}.
\newblock
\showISSN{1527-974X}
\urldef\tempurl%
\url{https://doi.org/10.1093/jamia/ocv094}
\showDOI{\tempurl}


\bibitem[Varanasi and Goyal(2023)]%
        {varanasi2023currently}
\bibfield{author}{\bibinfo{person}{Rama~Adithya Varanasi} {and} \bibinfo{person}{Nitesh Goyal}.} \bibinfo{year}{2023}\natexlab{}.
\newblock \showarticletitle{“It is currently hodgepodge”: Examining AI/ML Practitioners’ Challenges during Co-production of Responsible AI Values}. In \bibinfo{booktitle}{\emph{Proceedings of the 2023 CHI Conference on Human Factors in Computing Systems}}. \bibinfo{pages}{1--17}.
\newblock
\urldef\tempurl%
\url{https://doi.org/10.1145/3544548.3580903}
\showDOI{\tempurl}


\bibitem[Vines et~al\mbox{.}(2013)]%
        {Vines2013configuring}
\bibfield{author}{\bibinfo{person}{John Vines}, \bibinfo{person}{Rachel Clarke}, \bibinfo{person}{Peter Wright}, \bibinfo{person}{John McCarthy}, {and} \bibinfo{person}{Patrick Olivier}.} \bibinfo{year}{2013}\natexlab{}.
\newblock \showarticletitle{Configuring participation: on how we involve people in design}. In \bibinfo{booktitle}{\emph{Proceedings of the SIGCHI Conference on Human Factors in Computing Systems}} \emph{(\bibinfo{series}{CHI ’13})}. \bibinfo{publisher}{Association for Computing Machinery}, \bibinfo{address}{New York, NY, USA}, \bibinfo{pages}{429–438}.
\newblock
\showISBNx{978-1-4503-1899-0}
\urldef\tempurl%
\url{https://doi.org/10.1145/2470654.2470716}
\showDOI{\tempurl}


\bibitem[Vinuesa et~al\mbox{.}(2020)]%
        {vinuesa_role_2020}
\bibfield{author}{\bibinfo{person}{Ricardo Vinuesa}, \bibinfo{person}{Hossein Azizpour}, \bibinfo{person}{Iolanda Leite}, \bibinfo{person}{Madeline Balaam}, \bibinfo{person}{Virginia Dignum}, \bibinfo{person}{Sami Domisch}, \bibinfo{person}{Anna Felländer}, \bibinfo{person}{Simone~Daniela Langhans}, \bibinfo{person}{Max Tegmark}, {and} \bibinfo{person}{Francesco Fuso~Nerini}.} \bibinfo{year}{2020}\natexlab{}.
\newblock \showarticletitle{The role of artificial intelligence in achieving the Sustainable Development Goals}.
\newblock \bibinfo{journal}{\emph{Nature Communications}} \bibinfo{volume}{11}, \bibinfo{number}{1} (\bibinfo{date}{Jan.} \bibinfo{year}{2020}), \bibinfo{pages}{233}.
\newblock
\showISSN{2041-1723}
\urldef\tempurl%
\url{https://doi.org/10.1038/s41467-019-14108-y}
\showDOI{\tempurl}


\bibitem[Voida et~al\mbox{.}(2011)]%
        {voida2011homebrew}
\bibfield{author}{\bibinfo{person}{Amy Voida}, \bibinfo{person}{Ellie Harmon}, {and} \bibinfo{person}{Ban Al-Ani}.} \bibinfo{year}{2011}\natexlab{}.
\newblock \showarticletitle{Homebrew databases: Complexities of everyday information management in nonprofit organizations}. In \bibinfo{booktitle}{\emph{Proceedings of the SIGCHI Conference on Human Factors in Computing Systems}}. \bibinfo{pages}{915--924}.
\newblock
\urldef\tempurl%
\url{https://doi.org/10.1145/1978942.1979078}
\showDOI{\tempurl}


\bibitem[Wallerstein and Duran(2006)]%
        {Wallerstein_Duran_2006}
\bibfield{author}{\bibinfo{person}{Nina~B. Wallerstein} {and} \bibinfo{person}{Bonnie Duran}.} \bibinfo{year}{2006}\natexlab{}.
\newblock \showarticletitle{Using community-based participatory research to address health disparities}.
\newblock \bibinfo{journal}{\emph{Health Promotion Practice}} \bibinfo{volume}{7}, \bibinfo{number}{3} (\bibinfo{date}{July} \bibinfo{year}{2006}), \bibinfo{pages}{312–323}.
\newblock
\showISSN{1524-8399}
\urldef\tempurl%
\url{https://doi.org/10.1177/1524839906289376}
\showDOI{\tempurl}


\bibitem[Walsham(2017)]%
        {Walsham2017ictd}
\bibfield{author}{\bibinfo{person}{Geoff Walsham}.} \bibinfo{year}{2017}\natexlab{}.
\newblock \showarticletitle{ICT4D research: reflections on history and future agenda}.
\newblock \bibinfo{journal}{\emph{Information Technology for Development}} \bibinfo{volume}{23}, \bibinfo{number}{1} (\bibinfo{date}{Jan.} \bibinfo{year}{2017}), \bibinfo{pages}{18–41}.
\newblock
\showISSN{0268-1102}
\urldef\tempurl%
\url{https://doi.org/10.1080/02681102.2016.1246406}
\showDOI{\tempurl}


\bibitem[Wan et~al\mbox{.}(2023)]%
        {Wan_2023}
\bibfield{author}{\bibinfo{person}{Ruyuan Wan}, \bibinfo{person}{Adriana Alvarado~Garcia}, \bibinfo{person}{Devansh Saxena}, \bibinfo{person}{Catalina Vajiac}, \bibinfo{person}{Anna Kawakami}, \bibinfo{person}{Logan Stapleton}, \bibinfo{person}{Haiyi Zhu}, \bibinfo{person}{Kenneth Holstein}, \bibinfo{person}{Heloisa Candello}, {and} \bibinfo{person}{Karla Badillo-Urquiola}.} \bibinfo{year}{2023}\natexlab{}.
\newblock \showarticletitle{Community-driven AI: Empowering people through responsible data-driven decision-making}. In \bibinfo{booktitle}{\emph{Companion Publication of the 2023 Conference on Computer Supported Cooperative Work and Social Computing}} \emph{(\bibinfo{series}{CSCW ’23 Companion})}. \bibinfo{publisher}{Association for Computing Machinery}, \bibinfo{address}{New York, NY, USA}, \bibinfo{pages}{532–536}.
\newblock
\showISBNx{9798400701290}
\urldef\tempurl%
\url{https://doi.org/10.1145/3584931.3611282}
\showDOI{\tempurl}


\bibitem[Wang et~al\mbox{.}(2022)]%
        {wang2022whose}
\bibfield{author}{\bibinfo{person}{Ding Wang}, \bibinfo{person}{Shantanu Prabhat}, {and} \bibinfo{person}{Nithya Sambasivan}.} \bibinfo{year}{2022}\natexlab{}.
\newblock \showarticletitle{Whose AI Dream? In search of the aspiration in data annotation.}. In \bibinfo{booktitle}{\emph{Proceedings of the 2022 CHI Conference on Human Factors in Computing Systems}}. \bibinfo{pages}{1--16}.
\newblock


\bibitem[Winschiers-Theophilus et~al\mbox{.}(2010)]%
        {Winschiers-Theophilus_2010}
\bibfield{author}{\bibinfo{person}{Heike Winschiers-Theophilus}, \bibinfo{person}{Shilumbe Chivuno-Kuria}, \bibinfo{person}{Gereon~Koch Kapuire}, \bibinfo{person}{Nicola~J. Bidwell}, {and} \bibinfo{person}{Edwin Blake}.} \bibinfo{year}{2010}\natexlab{}.
\newblock \showarticletitle{Being participated: a community approach}. In \bibinfo{booktitle}{\emph{Proceedings of the 11th Biennial Participatory Design Conference}}. \bibinfo{publisher}{ACM}, \bibinfo{address}{Sydney Australia}, \bibinfo{pages}{1–10}.
\newblock
\showISBNx{978-1-4503-0131-2}
\urldef\tempurl%
\url{https://doi.org/10.1145/1900441.1900443}
\showDOI{\tempurl}


\bibitem[Wong et~al\mbox{.}(2023)]%
        {wong2023seeing}
\bibfield{author}{\bibinfo{person}{Richmond~Y Wong}, \bibinfo{person}{Michael~A Madaio}, {and} \bibinfo{person}{Nick Merrill}.} \bibinfo{year}{2023}\natexlab{}.
\newblock \showarticletitle{Seeing like a toolkit: How toolkits envision the work of AI ethics}.
\newblock \bibinfo{journal}{\emph{Proceedings of the ACM on Human-Computer Interaction}} \bibinfo{volume}{7}, \bibinfo{number}{CSCW1} (\bibinfo{year}{2023}), \bibinfo{pages}{1--27}.
\newblock


\bibitem[Yildirim et~al\mbox{.}(2023)]%
        {yildirim2023investigating}
\bibfield{author}{\bibinfo{person}{Nur Yildirim}, \bibinfo{person}{Mahima Pushkarna}, \bibinfo{person}{Nitesh Goyal}, \bibinfo{person}{Martin Wattenberg}, {and} \bibinfo{person}{Fernanda Vi{\'e}gas}.} \bibinfo{year}{2023}\natexlab{}.
\newblock \showarticletitle{Investigating How Practitioners Use Human-AI Guidelines: A Case Study on the People+ AI Guidebook}. In \bibinfo{booktitle}{\emph{Proceedings of the 2023 CHI Conference on Human Factors in Computing Systems}}. \bibinfo{pages}{1--13}.
\newblock
\urldef\tempurl%
\url{https://doi.org/10.1145/3544548.3580900}
\showDOI{\tempurl}


\bibitem[Yoon and Copeland(2020)]%
        {yoon2020toward}
\bibfield{author}{\bibinfo{person}{Ayoung Yoon} {and} \bibinfo{person}{Andrea Copeland}.} \bibinfo{year}{2020}\natexlab{}.
\newblock \showarticletitle{Toward community-inclusive data ecosystems: Challenges and opportunities of open data for community-based organizations}.
\newblock \bibinfo{journal}{\emph{Journal of the Association for Information Science and Technology}} \bibinfo{volume}{71}, \bibinfo{number}{12} (\bibinfo{year}{2020}), \bibinfo{pages}{1439--1454}.
\newblock
\urldef\tempurl%
\url{https://doi.org/10.1002/asi.24346}
\showDOI{\tempurl}


\bibitem[Zegura et~al\mbox{.}(2018)]%
        {zegura2018care}
\bibfield{author}{\bibinfo{person}{Ellen Zegura}, \bibinfo{person}{Carl DiSalvo}, {and} \bibinfo{person}{Amanda Meng}.} \bibinfo{year}{2018}\natexlab{}.
\newblock \showarticletitle{Care and the practice of data science for social good}. In \bibinfo{booktitle}{\emph{Proceedings of the 1st ACM SIGCAS Conference on Computing and Sustainable Societies}}. \bibinfo{pages}{1--9}.
\newblock
\urldef\tempurl%
\url{https://doi.org/10.1145/3209811.3209877}
\showDOI{\tempurl}


\bibitem[Zytko et~al\mbox{.}(2022)]%
        {Zytko_2022}
\bibfield{author}{\bibinfo{person}{Douglas Zytko}, \bibinfo{person}{Pamela J.~Wisniewski}, \bibinfo{person}{Shion Guha}, \bibinfo{person}{Eric P.~S.~Baumer}, {and} \bibinfo{person}{Min~Kyung Lee}.} \bibinfo{year}{2022}\natexlab{}.
\newblock \showarticletitle{Participatory Design of AI Systems: Opportunities and Challenges Across Diverse Users, Relationships, and Application Domains} \emph{(\bibinfo{series}{CHI EA ’22})}. \bibinfo{address}{New York, NY, USA}, \bibinfo{pages}{1–4}.
\newblock
\showISBNx{978-1-4503-9156-6}
\urldef\tempurl%
\url{https://doi.org/10.1145/3491101.3516506}
\showDOI{\tempurl}


\end{thebibliography}
